\documentclass[a4paper,fleqn,usenatbib,useAMS]{mnras}

\usepackage{newtxtext,newtxmath}

\usepackage[T1]{fontenc}
\usepackage{ae,aecompl}

\usepackage{graphicx}	
\usepackage{amsmath}	
\usepackage{mathtools}

\title[Wide-field LOFAR-LBA power-spectra analyses]{Wide-field LOFAR-LBA power-spectra analyses: Impact of calibration, polarization leakage and ionosphere}

\author[B. K. Gehlot et al.]{B. K. Gehlot$^{1}$\thanks{E-mail: gehlot@astro.rug.nl (BKG)}, L. V. E. Koopmans$^{1}$, A. G. de Bruyn$^{1,2}$\thanks{Deceased (July 9, 2017)}, S. Zaroubi$^{1,3}$,
\newauthor M. A. Brentjens$^{2}$, K. M. B. Asad$^{5,6,7,1}$, M. Hatef$^{1,2}$, V. Jeli\'c $^{4,2}$, M. Mevius$^{1,2}$,
\newauthor A. R. Offringa$^{2}$, V. N. Pandey$^{1,2}$, and S. Yatawatta$^{1,2}$
\\
$^{1}$Kapteyn Astronomical Institute, University of Groningen, PO Box 800, 9700AV Groningen, the Netherlands\\
$^{2}$ASTRON, PO Box 2, 7990AA Dwingeloo, The Netherlands\\
$^{3}$Department of Natural Sciences, The Open University of Israel, 1 University Road, PO Box 808, Ra'anana 4353701, Israel \\
$^{4}$Ru{\dj}er Bo\v{s}kovi\'{c} Institute, Bijeni\v{c}ka cesta 54, 10000 Zagreb, Croatia\\
$^{5}$Department of Physics, University of the Western Cape, Cape Town 7535, South Africa \\
$^{6}$Department of Physics and Electronics, Rhodes University, PO
Box 94, Grahamstown, 6140, South Africa\\
$^{7}$SKA South Africa, 3rd Floor, The Park, Park Road, Pinelands, 7405 South Africa
}

\date{Accepted XXX. Received YYY; in original form ZZZ}

\pubyear{2017}

\begin{document}
\label{firstpage}
\pagerange{\pageref{firstpage}--\pageref{lastpage}}
\maketitle

\begin{abstract}
Contamination due to foregrounds (Galactic and Extra-galactic), calibration errors and ionospheric effects pose major challenges in detection of the cosmic 21 cm signal in various Epoch of Reionization (EoR) experiments. We present the results of a pilot study of a field centered on 3C196 using LOFAR Low Band (56-70 MHz) observations, where we quantify various wide field and calibration effects such as gain errors, polarized foregrounds, and ionospheric effects. We observe a `pitchfork' structure in the 2D power spectrum of the polarized intensity in delay-baseline space, which leaks into the modes beyond the instrumental horizon (EoR/CD window). We show that this structure largely arises due to strong instrumental polarization leakage ($\sim30\%$) towards {Cas\,A} ($\sim21$ kJy at 81 MHz, brightest source in northern sky), which is far away from primary field of view. We measure an extremely small ionospheric diffractive scale ($r_{\text{diff}} \approx 430$ m at 60 MHz) towards {Cas\,A} resembling pure Kolmogorov turbulence compared to $r_{\text{diff}} \sim3 - 20$ km towards zenith at 150 MHz for typical ionospheric conditions. This is one of the smallest diffractive scales ever measured at these frequencies. Our work provides insights in understanding the nature of aforementioned effects and mitigating them in future Cosmic Dawn observations (e.g. with SKA-low and HERA) in the same frequency window. 
\end{abstract}

\begin{keywords}
techniques: interferometric -- dark ages, reionization, first stars -- polarization -- techniques: polarimetric -- atmospheric effects -- methods: statistical
\end{keywords}


\vspace{1cm}

\section{Introduction}
The first stars and galaxies formed during the so-called Cosmic Dawn (CD) spanning redshifts $30 \gtrsim z \gtrsim 15$ \citep{pritchard2007}. The ultraviolet and X-ray radiation from these first stars started to heat and ionize the neutral Hydrogen (HI hereafter) in the surrounding Inter-Galactic Medium (IGM), continuing until hydrogen gas in the universe transitioned from being fully neutral to become fully ionized \citep{madau1997}. Substantial ionization of the IGM only occurred at $z \lesssim 15$ and this process completed around $z \sim 6$. This era in the history of the universe is known as the Epoch of Reionization (EoR). 

Current constraints on the redshift range of the reionization are inferred from indirect probes such as high-redshift quasar spectra \citep{becker2001,fan2003,fan2006}, the optical depth for Thomson scattering from Cosmic Microwave Background (CMB) polarization anisotropy \citep{page2007,komatsu2011,hinshaw2013, planck2016}, IGM temperature measurements \citep{theuns2002,bolton2010}, Lyman break galaxies \citep{pentericci2011,ono2012,schenker2012}, the kinetic Sunyaev-Zel'dovich effect \citep{zahn2012} and high redshift gamma ray bursts \citep{wang2013}. 
The most recent constraint on the upper limit of reionization redshifts comes from \cite{planck2016} suggesting that the universe is ionized at $\lesssim 10 \%$ level for $z\gtrsim10$ and substantial reionization happened during redshifts between $z = 7.8$ and $8.8$. Although these probes shed some light on the timing and duration of the reionization,  there is very little known about the evolution of IGM during reionization, nature of sources of the ionizing radiation and their evolution. 

Observations of 21 cm hyperfine transition of HI at high redshifts promises to be an excellent probe of the  HI distribution in IGM during EoR \citep{madau1997, shaver1999, furlanetto2006, pritchard2012, zaroubi2013}. Several ongoing and upcoming experiments such as the LOw Frequency ARray\footnote{\url{http://www.lofar.org/}}(LOFAR; \citealt{vanhaarlem2013}), the Giant Meterwave Radio Telescope\footnote{\url{http://gmrt.ncra.tifr.res.in/}}(GMRT; \citealt{paciga2011}), the Murchison Widefield Array\footnote{\url{http://www.mwatelescope.org/}}(MWA; \citealt{tingay2013,bowman2013}), the Precision Array for Probing the Epoch of Reionization\footnote{\url{http://eor.berkeley.edu/}}(PAPER; \citealt{parsons2010}), the 21 Centimeter Array (21CMA; \citealt{zheng2016}), the Hydrogen Epoch of Reionization Array\footnote{\url{http://reionization.org/}}(HERA; \citealt{deboer2017}), and the Square Kilometer Array\footnote{\url{http://skatelescope.org/}}(SKA; \citealt{mellema2013,koopmans2015}) aim to detect the redshifted 21 cm emission from the EoR. Although the above instruments focus largely on detecting the EoR, LOFAR-LBA, and the upcoming HERA, SKA-low, LEDA\footnote{\url{http://www.tauceti.caltech.edu/leda/}}(Large-aperture Experiment to detect Dark Ages; \citealt{price2017}) and NENUFAR\footnote{\url{https://nenufar.obs-nancay.fr/}}(New Extension in Nan\c cay Upgrading loFAR; \citealt{zarka2012}) also observe at frequency range 50-80 MHz which corresponds to a part of the redshift range of the Cosmic Dawn ($30\gtrsim z \gtrsim 15$). In this paper, we focus on challenges for observing the Cosmic Dawn (CD) with LOFAR, and the future SKA-Low which will largely have a similar layout. Since these telescopes operate at a lower frequency range (50-80 MHz), they will face challenges (foregrounds and ionosphere) similar to EoR experiments but more severe in strength. 

The expected 21 cm signal from $z = 30 $ to $15$ is extremely faint with $\Delta_{21}^2 \sim (5-6 \ \text{mK})^2$ \citep{furlanetto2006,pritchard2012}. This signal is buried deep below galactic and extra-galactic foreground emission which dominate the sky at these low frequencies (50-80 MHz). The foreground emission is $\sim 4$ orders of magnitude larger in strength than the 21 cm signal and has a brightness temperature of several Kelvins \citep{bernardi2010} (on relevant angular scales) at high Galactic latitudes where LOFAR-EoR observing fields are located. Even if the foregrounds are removed with great accuracy, the noise per voxel in the images cubes after hundreds of hours of integration will still be orders of magnitude higher than the expected signal. Therefore, the current experiments (both EoR and CD) are aiming for a statistical detection of the signal instead of directly mapping out HI in IGM at high redshifts. The LOFAR EoR Key Science Project (KSP) currently predominantly focuses on a statistical detection of the redshifted 21 cm signal from $z = 7 $ to $12$ (110-180 MHz) using LOFAR-High Band Antenna (HBA) observations and measure its power spectrum as a function of redshift \citep{patil2017}. Contamination due to the (polarized) foregrounds, ionospheric propagation effects and systematic biases (e.g. station-beam errors) pose considerable challenges in the detection of this signal. It is crucial to remove these bright foregrounds and mitigate other effects accurately in order to obtain a reliable (accurate and precise) estimate of the 21 cm power spectrum. This requires a detailed understanding of the nature of these effects and the errors associated with these effects. Several contamination effects in LOFAR EoR observations (High Band Antenna, 110-180 MHz) have been studied in great detail, such as polarization leakage (see \citealt{asad2015,asad2016,asad2017}), systematic biases (see \citealt{patil2016}), ionospheric effects (see \citealt{vedantham2015,vedantham2016,mevius2016}), LOFAR Radio Frequency Interference (RFI) environment \citep{offringa2010,offringa2012,offringa2013a,offringa2013b}, calibration and effects of beam errors \citep{kazemi2011,kazemi2013a,kazemi2013b,yatawatta2013,yatawatta2015,yatawatta2016}. 

In this work, we study some of the aforementioned effects at low frequencies using LOFAR Low Band Antenna (LBA) observations of a field centered on 3C196 (3C196 field hereafter) at lower frequency (56-70 MHz), covering part of the CD, where both the foregrounds and ionospheric effects are known to be even stronger. LOFAR-HBA observations of the 3C196 field show bright polarized emission of $\sim$ few Kelvins with complicated and rich morphology \citep{jelic2015}. We address the broadband nature of the excess noise due to systematic biases, polarized foregrounds and ionospheric effects. A similar analysis has been done by \citealt{ewall-wice2016} using low-frequency MWA observations (75-112 MHz), which addresses the MWA RFI environment, instrumental, and ionospheric effects at these frequencies. Our analysis provides improved insight on the spectral behavior of the associated errors as well as the level of these contamination effects in ongoing and upcoming experiments to detect the HI signal from the CD era at low frequencies (50-80 MHz). 

The paper is organized as follows: in section 2, we briefly describe the data processing steps. In section 3, we discuss the differential Stokes power spectrum method to study excess noise and its behavior for different calibration strategies. In section 4, we discuss the delay power spectrum method to study the polarized foregrounds and polarization leakage. We also discuss the effect of different calibration strategies and source subtraction on polarization leakage. In section 5, we discuss the ionospheric effects at low frequencies using cross coherence method. In section 6, we provide conclusions and summary of the analysis in this work.

\section{Observations and Data processing}
\begin{figure*}
	\includegraphics[width=0.9\textwidth]{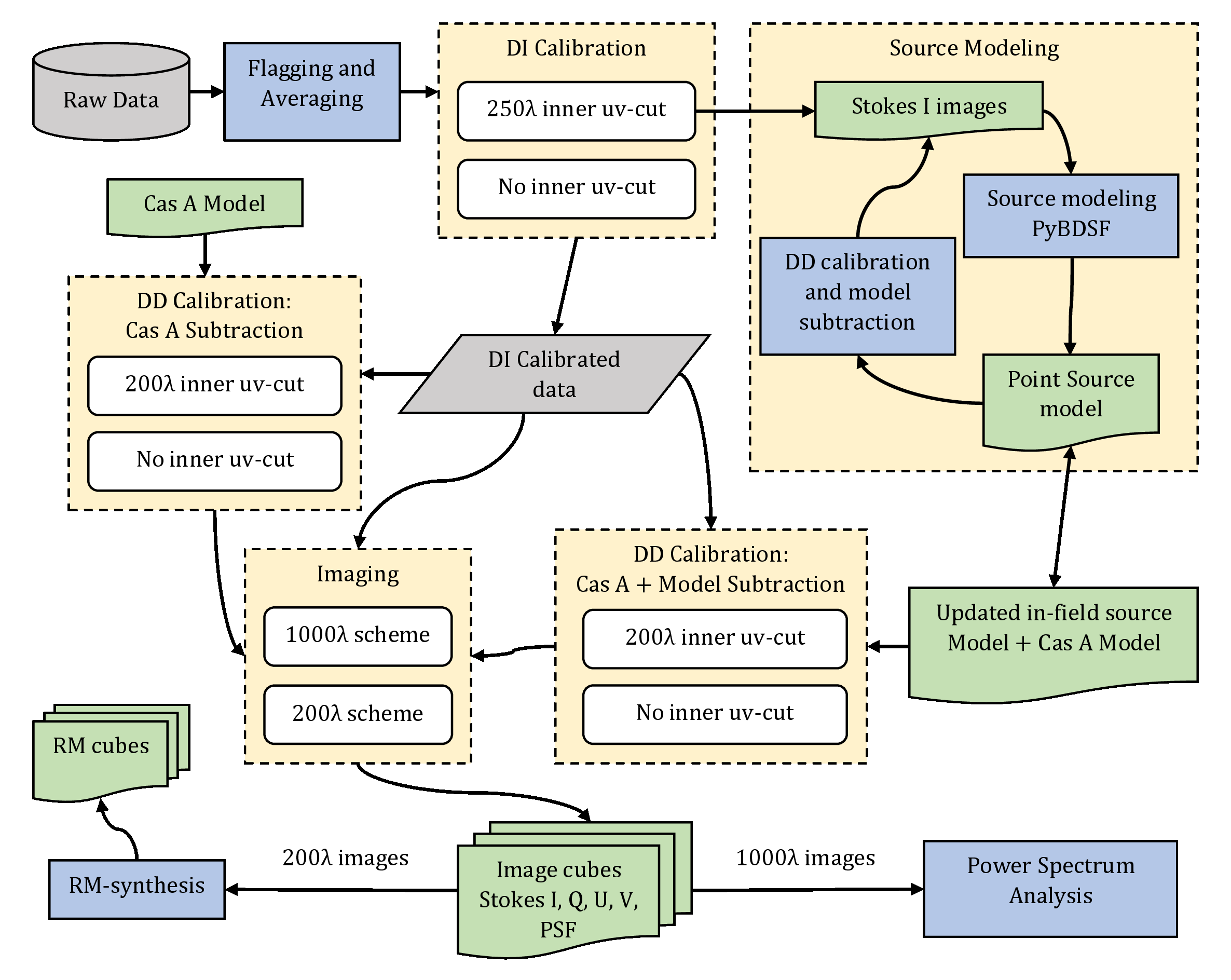}
    \caption{Flow chart illustrating the steps involved in data processing. Solid rectangles represent the processes. Rounded rectangles represent various schemes within a process. Dotted rectangles represent processes with sub-processes and/or multiple processing schemes. Parallelograms represent the stored visibility data. Boxes with curved bottom represent data products of the pipeline, such as sky-model and image cubes. Arrows represent the data flow.}
    \label{fig:flow_chart}
\end{figure*}

We have used LOFAR-LBA observations of the 3C196 field for our analysis, it being one of the two primary observation windows of the LOFAR EoR KSP. 3C196 is a relatively compact (4 arcsec) bright radio source placed at the center of the field and serves as a band-pass calibrator. Observed data was processed using the standard LOFAR software pipeline (see e.g. LOFAR imaging cookbook \footnote{\url{https://www.astron.nl/radio-observatory/lofar/lofar-imaging-cookbook}}). The observational setup and the steps for data processing are briefly described in the following subsections. Figure \ref{fig:flow_chart} shows a flow chart of the data processing steps.

\begin{table}
	\centering
	\caption{Observational details of the data.}
	\label{tab:obs_details}
	\begin{tabular}{ll} 
		\hline		
		\textbf{Parameter} & \textbf{value} \\	
		\hline
		Telescope & LOFAR LBA \\
		Observation cycle and ID & Cycle 0, L99269 \\ 		
		Antenna configuration & \texttt{LBA\_INNER} \\
		Number of stations & 37 (NL stations) \\
		Observation start time (UTC) & March 2, 2013;17:02:52 \\ 
		Phase center ($\alpha,\delta$; J2000) & 08h13m36s, $+48^{\circ}13^{\prime}03^{\prime\prime}$ \\
		Duration of observation & 8 hours   \\
		Frequency range & 30-78 MHz         \\
		Primary beam FWHM (at 60 MHz)& $9.77^{\circ}$ \\
		Field of View (at 60 MHz) & 75 $\text{deg}^2$ \\
		SEFD (at 60 MHz) & $\sim26$ kJy \\		
		Polarization & Linear X-Y   \\
		Time, frequency resolution: \\
		\quad Raw Data & 1 s, 3 kHz      \\
		\quad After flagging and averaging & 5 s, 183.1 kHz \\ 	
		\hline
	\end{tabular}
\end{table}

\subsection{LOFAR-LBA system}

The LOFAR array has 38 stations in the Netherlands, out of which 24 stations (also known as core stations) are spread within a core of 2 km radius, and 14 stations (known as remote stations) are spread across 40 km east-west and 80 km north-south area in northeastern part of the Netherlands. Each LOFAR station has 96 low band dual-polarization dipole antennas spread within an area of 87 m diameter. LBA dipoles have an arm length of 1.38 meter, which corresponds to a resonance frequency of 52 MHz. LBAs are designed to operate in the frequency range of 10-90 MHz, but the operational bandwidth of LBA is limited to 30-80 MHz to avoid strong RFI below 30 MHz and RFI due to proximity to the FM band above 80 MHz. At a given time, signals from only 48 out of 96 LBA dipoles can be processed. The signals from these 48 dipoles are digitized and beam-formed to produce a station beam which is steered digitally to track a fixed phase center in the sky. The LOFAR-LBA system offers three different LBA dipole configurations viz: \texttt{LBA\_INNER} where 48 innermost dipoles (array width $\sim 30$ m) are beam-formed, \texttt{LBA\_OUTER} where 48 outermost dipoles (array width $\sim 87$ m) are beam-formed, and \texttt{LBA\_SPARSE} where half of the innermost 48 dipoles, plus half of the outermost 48 dipoles (array width $\sim 87$ m) are beam-formed. These different configurations provide different Field of View (FoV) areas as well as different sensitivities due to mutual coupling between the dipoles. The data is digitized by the receivers with 200 MHz sampling clock, providing a RF bandwidth of 96 MHz. The digitized data is transported to the GPU correlator via a fiber optics network. The correlator generates visibilities with 3 kHz frequency resolution (64 channels per sub-band) and 1 s integration and stores them in a Measurement Set (MS) format. Readers may refer to \cite{vanhaarlem2013} for more information about LOFAR capabilities. 

\subsection{Observations}
We use 8 hours of synthesis observation data (L99269 (LOFAR Cycle 0): March 2-3, 2013) of 3C196 field (pointing/phase center: RA=08h13m36s, Dec=$+48^{\circ}13^{\prime}03^{\prime\prime}$, Epoch=J2000) using the LOFAR LBA system. The field was observed with 37 LOFAR-LBA stations in the Netherlands (70 m to 80 km baseline) operating in the frequency range of 30-78 MHz. The correlations of voltages from antenna pairs were recorded with 1 second time resolution and 3 kHz frequency resolution. The recorded data consists of 248 sub-bands, and each sub-band has 195.3 kHz width and consists of 64 channels. We used only 56-70 MHz band in our analysis, which is the most sensitive region of the LBA band and is relatively free from Radio Frequency Interference (RFI). Four out of eight observation hours are used in our analysis and we discarded the visibilities for the first two hours and last two hours of observation. The choice of this `hard cut' is based on the quality of station based gain solutions after direction independent calibration step. We observed that the phases of the gain solutions were varying rapidly as a function of time in the beginning and at the end of the observations. The rapid variation of phases of gain solutions represents strong ionospheric activity which leads to strong amplitude scintillation. The observational details of the data are summarized in Table \ref{tab:obs_details}. 

\subsection{Flagging and averaging}
The first step of the processing is flagging of RFI-corrupted data. RFI mitigation usually works best on the highest resolution data in order to minimize any information loss. Thus, this step is performed on the raw data with the time and frequency resolution of 1 s and 3 kHz. RFI mitigation is performed using the \texttt{AOFlagger} software \citep{offringa2010,offringa2012}. Two channels on either edge of every sub-band are discarded in order to avoid edge effects due to the polyphase filter, resulting in a final width of 183.1 kHz per sub-band. Note that the separation between two consecutive sub-bands is still 195.3 kHz. After flagging, the remaining data are averaged to 5 second and to 183.1 kHz sub-band resolution. These resolutions are chosen such that the time and frequency smearing is limited to the longer baselines and does not affect the baselines of interest ($\lesssim 1000 \lambda$). In addition, we flagged 3 stations CS013LBA, CS030LBA and RS409LBA, which have 4, 6 and 10 non-working dipoles respectively, on the basis of their poor quality of the direction independent gain solutions. 

\begin{table*}
	\centering
	\caption{Calibration Parameters}
	\label{tab:cal_params}
	\begin{tabular}{lll} 
		\hline
		\multicolumn{3}{|c|}{Direction Independent calibration} \\
		\hline		
		\textbf{Parameter} & \textbf{Value} & \textbf{comments} \\
		\hline
		Flux Calibrator & 3C196 & J2000: 08h13m36s, $+48^{\circ}13^{\prime}03^{\prime}$\\ 
		Sky-model components & 4 & Gaussian \\
		Source spectral order ($n$) & 2 & log-polynomial spectra; $\log P_n = \log S_{\circ} + {\sum}_{i=1}^{n} \alpha_{i} \left[ \log \left( \dfrac{\nu}{\nu_{\circ}} \right) \right]^n$ \\
		Calibration baselines & 1. $\geq250\lambda$ & Two strategies \\
		 & 2. All ($\geq 0 \lambda$) \\
		
		Solution type & Full Jones & solves for all polarizations \\
		Solution interval: \\ 
		\quad time & 30 seconds \\
		\quad frequency & 183.1 kHz \\		
		\hline
		\multicolumn{3}{|c|}{Direction Dependent calibration} \\	
		\hline		
		Sky-model components & 188 & Compact; with apparent fluxes\\
		{Cas\,A} model components & 25 & 11 Gaussian + 14 compact; with apparent fluxes \\
		Source spectral order ($n$) & 1 & log-polynomial spectra; $\log P_n = \log S_{\circ} + {\sum}_{i=1}^{n} \alpha_{i} \left[ \log \left( \dfrac{\nu}{\nu_{\circ}} \right) \right]^n$\\		
		Calibration directions & 5 & 4 within FoV and 1 on {Cas\,A} \\			 
		Calibration baselines & 1. $\geq200\lambda$ & Two strategies \\
		 & 2. All ($\geq 0 \lambda$) \\
		 
		Solution type & Full Jones & solves for all polarizations \\
		Solution interval: \\ 
		\quad time & 5 minutes \\
		\quad frequency & 183.1 kHz \\
		\hline 
	\end{tabular}
\end{table*}

\subsection{Calibration}
The sky observed by LOFAR is distorted by the characteristics of the instrument (station beam, global band pass, clock drift etc.) and the environment (ionosphere). Calibration of a radio telescope refers to the estimation of the errors that corrupt the visibilities measured by the telescope, and to obtain an accurate estimate of the visibilities from the observed data. The influence of the instrument and the environment on the measured visibilities can be described by the radio interferometer \textit{measurement equation} \citep{hamaker1996,smirnov2011a,smirnov2011b}. The effects that corrupt the observed visibilities can be divided into two categories: (a) \textit{Direction Independent effects} (DIEs) and, (b) \textit{Direction Dependent Effects} (DDEs). 

DIEs are instrument related effects which are independent of the sky direction. These include complex antenna gains and frequency band-pass, as well as, a single phase and amplitude correction for the average ionosphere above each station. The DDEs vary as a function of the sky direction. These are, for example, caused by antenna voltage patterns, ionospheric phase fluctuations and Faraday rotation.  

\subsubsection{Direction Independent Calibration}  
Direction independent calibration refers to the estimation of a single instrumental gain for each beam-formed interferometric element (a station-in the context of LOFAR). LOFAR station gain is described by a complex $2 \times 2$ Jones matrix and represents two linear polarizations. The QSO 3C196 is a very bright radio source with known flux (130 Jy at 74 MHz; \citealt{kassim2007}) and is located at the phase center of the field. It can be used as a flux calibrator to determine the station band-pass gains. We use a model of 3C196\footnote{V. N. Pandey via private communication} which has 4 Gaussian components to describe the source, and the source spectrum is described by a second-order log-polynomial. This model was iteratively derived using LOFAR-HBA (full Dutch array with baseline range of 100 m to 120 km) observation data of 3C196 over the frequency range of 115-185 MHz. The parameters of the model components including the spectral indices were obtained by fitting in the visibility domain. The source model includes the flux at large angular scales and also represents the high resolution structures (with arcsec accuracy). We compared the 3C196 model flux extrapolated at lower frequencies with other 3C196 observations at 74 MHz using Very Large Array (VLA) \citep{kassim2007} and at 60 MHz using Serpukov radio-telescope \citep{aslanian1968}. The model flux matches the VLA observation within $4\%$ error and the Serpukov radio-telescope observation within $2\%$ error. Hence, the model performs well at the frequencies of interest. We use the Black Board Selfcal (BBS) package \citep{pandey2009} to obtain and subsequently apply the gain solutions for 30 s and 183.1 kHz intervals. 3C196 is subtracted in this step, and the residual visibilities are used for further processing. We use two different strategies for DI calibration: (1) using the baselines which are $\geq250\lambda$ for calibration (``$250 \lambda$ cut", hereafter), to avoid model incompleteness due to diffuse emission (see \citealt{patil2016,patil2017}), and (2) using all baselines (``no cut", hereafter) for calibration. The reasoning behind this is to reflect the effect of including/excluding small baselines and inclusion/exclusion of unmodeled diffuse flux on the calibration products. This is further explained in later sections. Parameters for the DI calibration steps are listed in Table \ref{tab:cal_params}.

\begin{figure*}
	\includegraphics[width=1.0\textwidth]{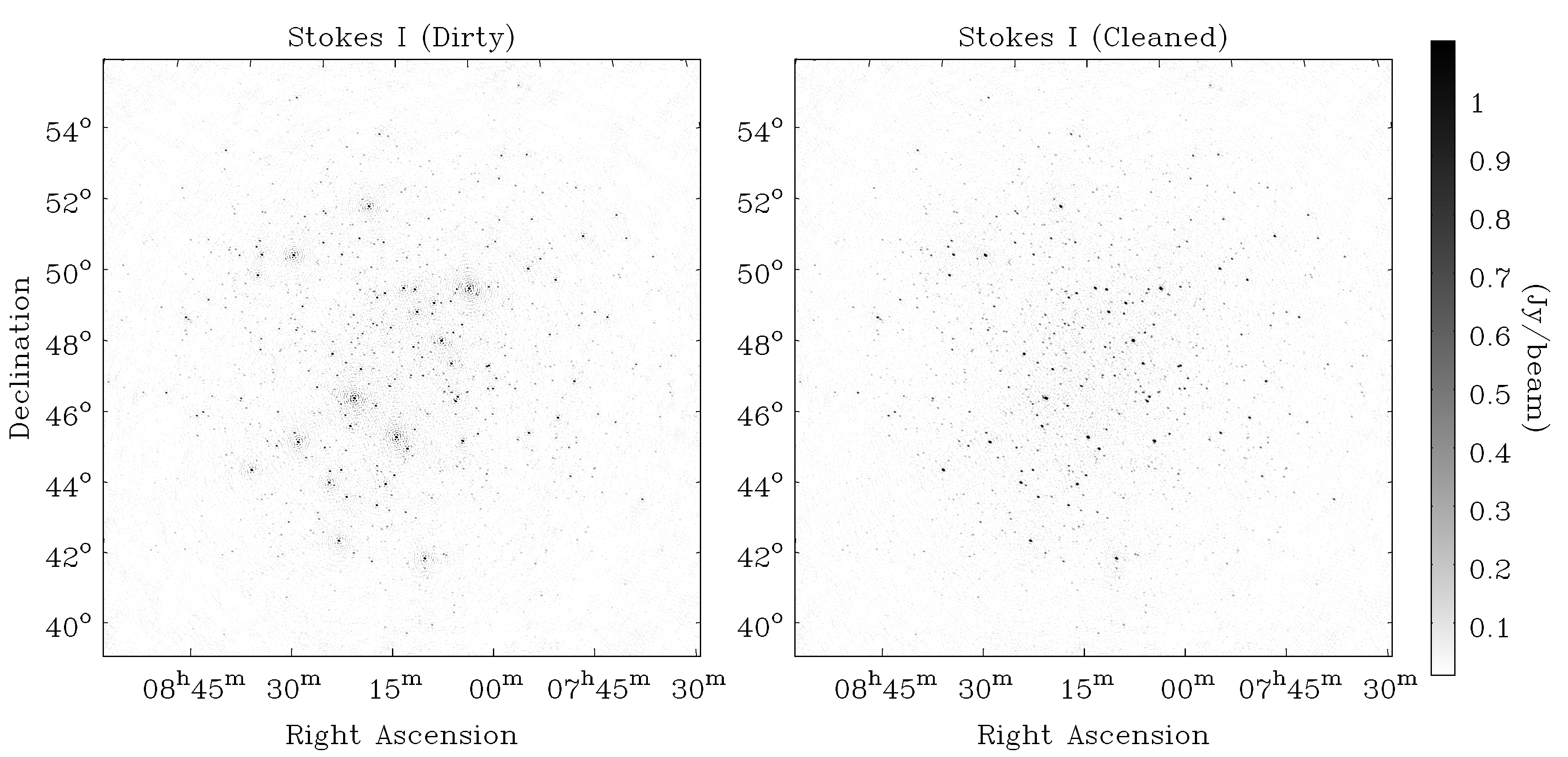}
    \caption{Direction Independent calibrated continuum images (56-70 MHz) of the 3C196 field. The left panel shows the dirty image and right panel shows the cleaned image. Calibration was done using $250\lambda$ cut parameters and imaging using $\leq1000\lambda$ baselines with uniform weights. The image has a point source rms $\sigma_{\text{image}}\sim 27$ mJy, whereas the expected theoretical value of the thermal noise is $\sigma_{\text{th}} \sim 2.1$ mJy for the observation, which is around 13 times smaller than the observed value. $\sigma_{\text{th}}$ can be calculated as $\sigma_{\text{th}} =  \text{SEFD}/\sqrt{N(N-1)\Delta \nu \Delta t}$, where SEFD (System Equivalent Flux Density) $\sim 26$ kJy at 60 MHz, $N = 30$ (corresponding to $1000\lambda$ imaging cut), $\Delta\nu = 13.7$ MHz, $\Delta t = 0.9\times 4$ hours (assuming flagged data at $10\%$ level).}  
\label{fig:cont_image}    
\end{figure*}

\subsubsection{Direction Dependent Calibration and source subtraction}
The low-frequency radio sky is dominated by galactic diffuse foregrounds (synchrotron, free-free emission) and extra-galactic compact sources (radio galaxies, supernova remnants). The Galactic diffuse emission at high Galactic latitudes dominates only on small baselines ($\leq 50 \lambda$) and LOFAR-LBA has very few baselines $\leq 50 \lambda$ at lower frequencies causing lesser sensitivity. Hence, the diffuse emission is mostly undetectable in the images. LOFAR-LBA images are dominated by the extra-galactic compact sources which need to be removed in order to obtain a clean power spectrum relatively free from the foregrounds. The signal arriving from different directions, however, is corrupted by direction dependent errors, which arise from wave propagation effects through the ionosphere and the primary beam (i.e. gain errors per station receiver element). These effects can produce artifacts around bright sources making it difficult to subtract them without leaving strong artifacts in the images. These effects can be accounted for during source subtraction by using Direction Dependent (DD) calibration. This requires obtaining the gain solutions in multiple directions. We use \texttt{SAGECal} \citep{kazemi2011,kazemi2013a,kazemi2013b,yatawatta2015,yatawatta2016} for direction dependent calibration and source subtraction. Note that we do not perform consensus optimization (\texttt{SAGECal-CO} which is a more recent addition to \texttt{SAGECal}) while solving for the gains, but solve for each sub-band independently from the other sub-bands. The sources in the calibration model are removed by multiplying the obtained gain solutions with the predicted visibilities and subtracting the product from the observed visibilities. In the DD-calibration step, we provide a sky-model consisting of 188 compact sources within the primary beam FoV (in-field model, hereafter) with a flux density range between 300 mJy to 11 Jy (described in later section) and {Cas\,A} (25 components\footnote{{Cas\,A} model is derived from a single sub-band {Cas\,A} image (with 40 arcsec restored beam size) produced using LOFAR-LBA at 52 MHz \citep{asgekar2013}. We used the source spectrum with spectral index of -0.77 \citep{baars1977}.}) containing positions, apparent fluxes, and spectral indices of the sources as an input for \texttt{SAGECal}. We solve for 5 directions; four directions are within the primary beam (each quadrant) and one is towards {Cas\,A}. Choosing only 4 directions within the primary beam optimizes signal-to-noise in each direction as well as minimizes the image noise which allows us to subtract more fainter sources compared to more number of directions. We choose the solution interval of 5 minutes and 183.1 kHz for the gain solutions. Subtracting {Cas\,A} is important, because the bright sources such as {Cas\,A} and {Cyg\,A} can cause significant sidelobe noise in the images even if these are far outside ($\gtrsim 40^{\circ}$) the primary beam. {Cyg\,A} ($\sim90^{\circ}$ away from the phase center) does not affect our observations because it is close to the horizon during the entire observation (discussed later). The residual visibilities after the DD calibration step and source subtraction are stored and imaged for further analysis. We also perform an alternative DD calibration step where we only subtract {Cas\,A} and image the residuals. In both cases, we choose two calibration strategies: (1) with $\geq 200\lambda$ baselines (``$200\lambda$ cut", hereafter) to avoid diffuse emission (absent in the sky-model) biasing the gain solutions and (2) using all baselines. Note that we use $\geq 200\lambda$ baselines in DD calibration instead using $\geq 250\lambda$ baselines as in DI calibration. We noticed that choosing $\geq 250\lambda$ cut in DD calibration produces noisy gain solutions across several sub-bands causing comparatively higher image rms values in these sub-bands. We think that the main reason behind these noisy gain solutions is the low Signal to Noise Ratio (SNR) in each direction which is solved for in DD calibration step. The SNR increases when we add more baselines to the calibration step by lowering the calibration cut to $\geq 200\lambda$ still without adding significant unmodeled diffuse emission in the calibration. The $\geq 200\lambda$ cut results in images with lower image rms values compared to the former. Parameters for the DD calibration steps are listed in Table \ref{tab:cal_params}.

\begin{table}
	\centering
	\caption{Imaging Parameters}
	\label{tab:img_params}
	\begin{tabular}{lll} 
		\hline 				
		\hline
		\textbf{Parameter} & \multicolumn{2}{|c|}{\textbf{value}} \\
		\hline	
		\hline		
		Imaging Scheme & $1000\lambda$ & $200\lambda$ \\		
		Imaging Baselines & 0-1000$\lambda$ & 0-200$\lambda$ \\		
		Frequency range & 56-70 MHz & 56-70 MHz	    \\					
		Weighting scheme & Uniform & Uniform        \\
		Spatial Resolution & 2.75 Arcmin & 13.75 Arcmin \\
		Pixel size & 45 Arcsec	& 3 Arcmin         \\   
		Number of Pixels & $1200 \times 1200$ & $192 \times 192$        \\
		\hline
	\end{tabular}
\end{table} 

\subsection{Imaging}
We use the \texttt{WSClean} \citep{offringa2014} package to image the visibilities. \texttt{WSClean} is a CPU-based imager and produces Stokes $I$, $Q$, $U$, $V$, and Point Spread Function (PSF) images as output. We image the visibilities after DI-calibration step and DD-calibration step for both calibration strategies. We use two different imaging schemes viz. $1000\lambda$ imaging and $200\lambda$ imaging. The $1000\lambda$ imaging scheme employs $0-1000\lambda$ baselines for imaging, and the output images are used for the power spectrum analysis. The $200\lambda$ imaging scheme uses $0-200\lambda$ baselines for imaging, and the output images are used to perform the Rotation Measure (RM) synthesis (see appendix \ref{appendix:rmsyn}). Both schemes use `uniform' weighting to achieve a cleaner side-lobe response. Although `natural' weighting scheme produces images with higher Signal to Noise Ratio (SNR) values compared to `uniform' weighting scheme, it produces a biased result in uv-space which has to be re-normalized to remove the effect of the gridding weights (i.e. tapering). Making a power spectrum from natural weighted images requires dividing the flux density in each $uv$-cell by its sampling density to get an unbiased power spectrum. It is mathematically almost equivalent to uniform weighting, which of course also performs this division. The only difference is when the ``kernels" (antialising, beam, etc.) are applied. Since the $uv$-coverage of LOFAR over the measured $uv$-cells is almost uniform, making power spectra from uniform and natural images results in similar power spectra. The reason we have used uniform weighting here is that uniform weighted images are easier to interpret and produce unbiased power spectra. The imaging parameters for both schemes are listed in Table \ref{tab:img_params}. Figure \ref{fig:cont_image} shows the DI calibrated (using $250\lambda$) dirty and cleaned Stokes $I$ continuum image (56-70 MHz) of 3C196 field where 3C196 has been subtracted off. {Cas\,A} has also been subtracted off using DD calibration (using $200\lambda$ scheme). We performed the multi-frequency deconvolution using \texttt{WSClean} with a cleaning threshold of 50 mJy.

\subsection{Source modeling}
Radio galaxies, galaxy clusters and supernova remnants are the discrete foreground sources observed at low radio frequencies. We used the Python Blob Detection and Source Finder (PyBDSF) software \citep{mohan2015} to model the bright compact sources in the 3C196 field. Source modeling is an iterative process where DI calibrated images are used to model the sources above a particular SNR threshold and determine their frequency spectra. The resulting sky-model from PyBDSF is used to perform DD calibration and source subtraction (see section 2.4.2) on the DI calibrated visibilities. The Stokes $I$ images of the residual visibilities are again modeled with PyBDSF to include fainter sources. This process is repeated until the confusion limit ($\sim90$ mJy at 60 MHz) is reached in a single sub-band. We have not applied any beam model\footnote{Beam model for LOFAR-LBA is a very recent addition to the LOFAR data processing pipeline and is still being improved. There were no beam models available for LBA when we performed most of the analysis. We only used the beam model in simulations (discussed later in section 4.1.3). The current version of the LOFAR-LBA beam model is derived from the Electro-Magnetic simulations of LOFAR-LBA dipoles. Beam model for LBA will be taken into account in future analyses.} to the data prior to modeling, which means that the modeled fluxes are apparent and averaged over the on-sky time. We create $15^{\circ}\times15^{\circ}$ images (centered around the primary beam) with pixel size of 45 arcsec using the $1000\lambda$ imaging scheme (see section 2.5 and Table \ref{tab:img_params}) for source modeling. We use a comb configuration with 12 sub-bands evenly spread across 56 MHz to 70 MHz for spectral index estimation. The final sky-model contains source positions, apparent fluxes and source spectra for 188 compact sources which have flux densities $\gtrsim 2.5$ times the rms noise in a single sub-band image. We use $1000 \lambda$ imaging scheme for the modeling purpose because the images produced with baselines greater than $1000 \lambda$ produces artifacts in the residual images after source subtraction as well as the PSF is more symmetric in $1000 \lambda$ images compared to the other cases.

\begin{figure*}
\centering
\includegraphics[width=\textwidth]{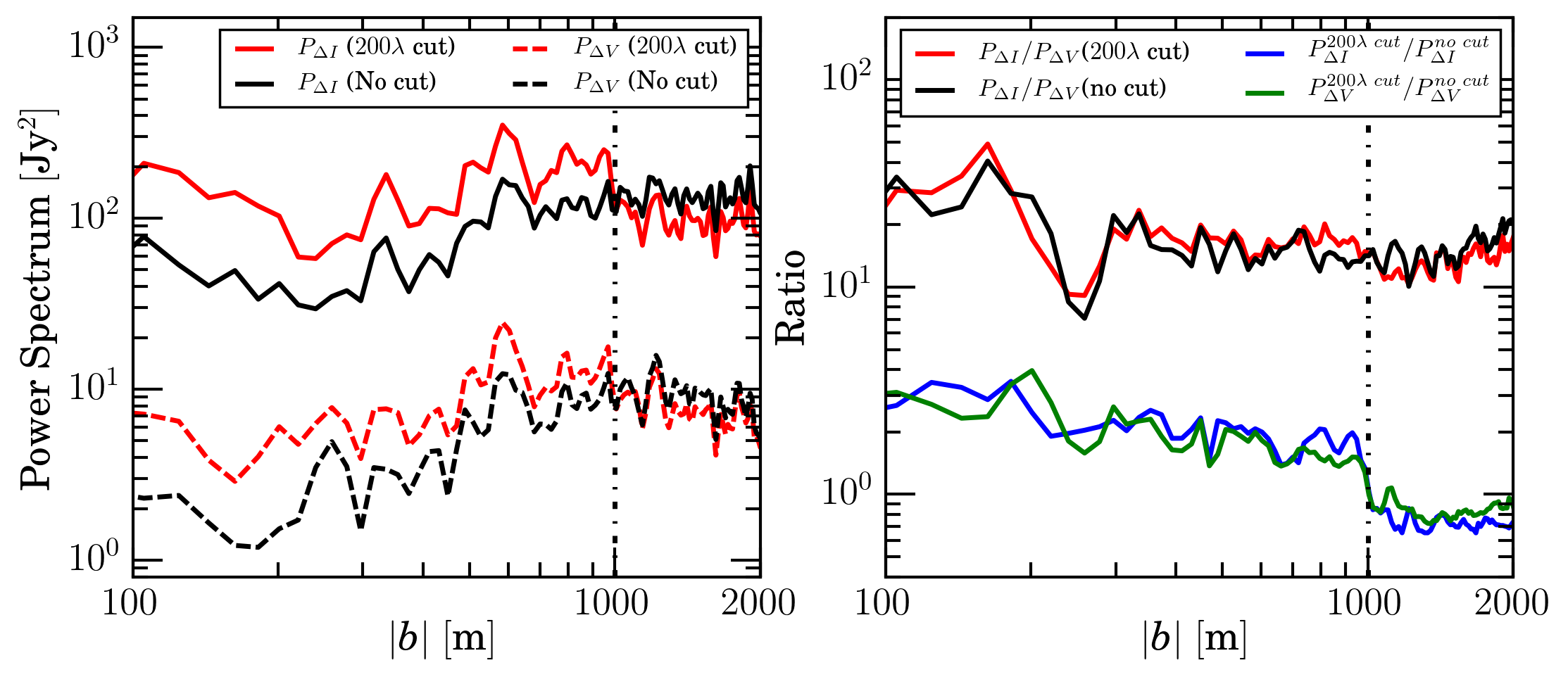}
    \caption{The left panel shows the differential Stokes $I$ and $V$ power spectra  for the two calibration schemes. Solid lines correspond to $P_{\Delta I}$ and dashed lines correspond to $P_{\Delta V}$. The color codes represent different calibration schemes; red color corresponds to $200\lambda$ cut strategy and black color corresponds all baselines (no cut strategy). Right panel shows the ratios of different combinations of $P_{\Delta I}$ and $P_{\Delta V}$. The red and black color correspond to the ratio $P_{\Delta I}/P_{\Delta V}$ for $200\lambda$ cut scheme and no cut respectively. The blue curve corresponds to the ratio $P_{\Delta I}(200\lambda \ \text{cut})/P_{\Delta I} (\text{No cut})$ and the green curve corresponds to the ratio $P_{\Delta V} (200\lambda\ \text{cut})/P_{\Delta V} (\text{No cut})$. The vertical dot-dashed line shows the location of the $200\lambda$ baseline at 60 MHz.}
\label{fig:dfferentialspectra}
\end{figure*}

\section{Differential power Spectrum}
Azimuthally averaged power spectrum of the difference between the Stokes images of adjacent sub-bands (differential Stokes images, hereafter) may be used to quantify the effects which are non-smooth in frequency (on sub-band level) such as instrumental and calibration effects. In an ideal scenario, total signal in a Stokes $I$ image at a given frequency can, to the first order, be expressed as a sum of the total sky signal convolved with the PSF and additive noise (see e.g. \citealt{patil2016}). Let $I_1$ and $I_2$ be the Stokes $I$ images at two consecutive frequency sub-bands, and $V_1$ and $V_2$ be the Stokes $V$ images respectively. We can write:
\begin{gather}
I_1 = S_1 \ast P_1 + N_1^I, \ \text{and} \ V_1 = N_1^V \\ 
I_2 = S_2 \ast P_2 + N_2^I, \ \text{and} \ V_2 = N_2^V
\end{gather}
where $S$ is the sky signal, $P$ is the PSF, $N^I$ and $N^V$ represent the noise in Stokes $I$ and $V$ images. We assume that the signal from the sky does not change within the 195 kHz frequency separation, which is the separation between two consecutive frequency sub-bands, i.e. $S = S_1 \approx S_2$. By making such assumption, we expect all the effects contributed by the foregrounds and ionosphere (assuming smoothness in frequency) to drop out, but the effects which are non-smooth in frequency on sub-band level are expected to remain. Therefore,
\begin{gather}
\Delta I = I_1 - I_2 = S \ast (P_1 - P_2) + (N_1^I - N_2^I) \\
\Delta V = V_1 - V_2 = (N_1^V - N_2^V)
\end{gather}
In Fourier space, equations 3 and 4 can be written as,
\begin{gather}
\tilde{\Delta I} = \tilde{S} \times \tilde{dP} + \tilde{N_1^I}-\tilde{N_2^I}  \\
\tilde{\Delta V} = \tilde{N_1^V} - \tilde{N_2^V}
\end{gather}
where the tilde represents Fourier transform (FT) and $dP = P_1 - P_2$ is the differential PSF due to slightly different $uv$-coverage. The spatial power-spectrum of the difference, $|\tilde{\Delta I}|^2$, is divided into $M$ annuli of width $\delta b = 19.1$ m in the $uv$-plane, and all the points within an annulus are averaged to obtain an estimate of the power. The differential power spectrum can finally be written as:   
\begin{gather}
P_{\Delta I} = \langle |\Delta \tilde{I}|^2 \rangle =  |\tilde{S}|^2 |\tilde{dP}|^2 + \langle |\tilde{N_1^I}|^2 \rangle + \langle |\tilde{N_2^I}|^2 \rangle \\
P_{\Delta V} = \langle |\Delta \tilde{V}|^2 \rangle = \langle |\tilde{N_1^V}|^2 \rangle + \langle |\tilde{N_2^V}|^2 \rangle
\end{gather}
where $P_{\Delta I}$ and $P_{\Delta V}$ represent azimuthally averaged Stokes $I$ and $V$ power spectra respectively. We use the DD calibrated residual images produced using the $1000\lambda$ imaging scheme with sub-band frequencies $\nu_1 =  59.7641$ MHz and $\nu_2 =  59.9594$ MHz to calculate $P_{\Delta I}$ and $P_{\Delta V}$. The selected sub-bands lie in the most sensitive region of the frequency band and are free from RFI. We estimate the power spectra for both calibration strategies.

\subsection{Excess Noise}
The sky signal has negligible circularly polarized component which is assumed to be well below the thermal noise. Because of this, Stokes $V$ can be used as a proxy for the thermal noise of the system. However, we observed that the point source rms value in the Stokes $V$ image for a single sub-band at 60 MHz is $\sigma_{\text{image}}\sim30$ mJy, which is $\sim 1.6$ times the theoretical value $\sigma_{\text{th}} \sim 18$ mJy (calculated using $\sigma_{\text{th}} = \text{SEFD}/\sqrt{N(N-1)\Delta \nu \Delta t}$, where SEFD $\sim 26$ kJy at 60 MHz, $N = 30$, $\Delta\nu = 183.1$ kHz, $\Delta t = 0.9\times 4$ hours). We think that this excess Stokes $V$ rms is due to the errors on the gain solutions which are applied to all polarizations during calibration step. Since each sub-band has different realizations of noise, the noise from two different sub-bands does not correlate. Also, the thermal noise in Stokes $I$ and $V$ is expected to be identical (see e.g. \citealt{vanstraten2009}), which means that they have identical statistical properties (e.g. variance). This leads us to define the excess noise ($P_{X}$) in Stokes $I$ as: 
\begin{equation}
P_{X} = P_{\Delta I} - P_{\Delta V} = \langle |\Delta \tilde{I}|^2 \rangle - \langle |\Delta \tilde{V}|^2 \rangle \ .  
\end{equation}
$P_{X}$ can be interpreted as excess power in differential Stokes $I$ compared to differential Stokes $V$. Figure \ref{fig:dfferentialspectra} shows $P_{\Delta I}$ and $P_{\Delta V}$ for the both $200\lambda$ cut and all baselines strategies. The right panel of figure 3 shows the ratio $P_{\Delta I}/P_{\Delta V} = P_{X}/P_{\Delta V} + 1$ for the both calibration strategies. We observe that $P_{X}$ is $\gtrsim 10$ times higher than $P_{\Delta V}$. Ideally, if the noise in Stokes $I$ and $V$ are statistically identical then, $P_{X} \approx |\tilde{S}|^2 |\tilde{dP}|^2$, which is the contribution due to chromatic PSF. Contribution due to chromatic PSF can be estimated by multiplying $|d\tilde{P}|^2$ with the sky contribution $|\tilde{S}|^2$ in Fourier space (readers may refer to \citealt{patil2016} for detailed calculation of chromatic PSF contribution). \cite{patil2016} showed that the contribution due to chromatic PSF in LOFAR-HBA observations is a small fraction of the excess noise. We also observe a similar behavior in LBA observations; the chromatic PSF seems to contribute less than $20\%$ to the overall excess noise between sub-bands on the relevant baselines.  The sky brightness also varies as a function of frequency (diffuse emission has $\nu^{-2.55}$ dependence), causing a brightness change of $\sim 1\%$ for 183.1 kHz difference between sub-bands. This is also a negligible effect compared to the excess noise we have observed, but it might become relevant in deeper experiments. We see a factor $\gtrsim 10$ larger power (i.e. $\gtrsim 3\times$ larger rms) in differential Stokes $I$ than Stokes $V$ for both calibration strategies. This ratio is almost constant as a function of baseline length and does not change between the two calibration strategies we employed. Introducing a calibration cut, however, decreases the power on baselines outside the cut and increases it on baselines inside the cut. However, the power in differential Stokes $I$ when calibrated using all baselines seems to decrease on smaller baselines. This decrease in power might occur because a diffuse sky-model is not included in the calibration.
There may be several causes of the significant excess power in $P_{\Delta I}$. These factors could include incomplete sky-model, imperfect source subtraction and ionospheric effects \citep{patil2016,barry2016,ewall-wice2017}. 

\subsection{Effect of calibration cut}
In the calibration scheme employed by \cite{patil2016,patil2017}, small baselines are excluded in calibration steps. The reasoning behind this is that small baselines ($<100 \lambda$) are dominated by diffuse foreground emission, and it is more difficult to model this emission and include it in the calibration model. One way to avoid any unmodeled flux biasing the calibration process is by choosing only those baselines where the diffuse emission is already resolved out. In such calibration schemes, longer baselines are used to obtain the gain solutions which are applied to all the visibilities, including shorter baselines. We compare the excess noise in the differential Stokes $I$ power spectrum in the two calibration strategies we employed. Figure \ref{fig:dfferentialspectra} shows the ratio $P_{\Delta I} (200\lambda \ \text{cut})/P_{\Delta I}(\text{no cut})$ and $P_{\Delta V} (200\lambda \ \text{cut})/P_{\Delta V}(\text{no cut})$. We observe that both ratios have a discontinuity at the exact location of the calibration cut. The excess noise, suddenly, is $\gtrsim 2$ times higher on baselines $< 200\lambda$ than on baselines $> 200\lambda$. This has also been observed in LOFAR HBA observations by \cite{patil2016}. This ratio is no longer constant on baselines $\leq 200\lambda$, but shows a slope with increasing excess power at shorter baselines. This effect is not only limited to Stokes $I$ but also present in Stokes $Q$, $U$ and $V$. We do not show the ratios for $Q$, $U$ here. We expect this effect to be purely because of the calibration cut. Because we perform a full Jones gain calibration, we expect this discontinuity to be present in all the Stokes parameters. Given that all Stokes power-spectra increase in the same manner, whereas their ratio with Stokes $V$ does not show any sign of change, suggests that this is the result of random errors introduced in the Jones matrices during the calibration process, which are subsequently applied to the sky-model and transferred to the image residuals during model subtraction. The cause of these random gain errors on the longer baselines could be due to sky-model incompleteness or the ionosphere \citep{patil2016,barry2016,ewall-wice2017}. Although differencing between sub-bands is a good first-order sanity check of whether the data reaches the expected noise-levels, a more powerful analysis can be carried out by using the combined information in all sub-bands. This is discussed in the next section.

\section{Delay Spectrum of gridded visibilities}
The delay spectrum is a powerful tool to study foregrounds and various contamination effects which can leak foregrounds into the EoR-window. A delay spectrum (see e.g. \citealt{parsons2009,parsons2012}) is defined as the FT of the visibilities along the frequency axis. Consider the gridded visibilities, $\mathcal{V}(u,v;\nu)$, as a function of baseline coordinates $(u,v)$\footnote{In radio interferometric imaging, the $(u,v)$ coordinates are defined in units of wavelength ($\lambda$) and are frequency invariant. Whereas, a delay spectrum is defined for baseline coordinates in physical units (meters) such that frequency dependence of baseline length is inherent to the delay transform.} and frequency $\nu$. Then:
\begin{gather}
\tilde{\mathcal{V}}_S(u,v;\tau) = \int_{-\infty}^{\infty} \mathcal{V}_S(u,v;\nu) e^{-2\pi i \nu \tau} d\nu \ ,\\
P_S(u,v;\tau) = |\tilde{\mathcal{V}}_S(u,v;\tau)|^2 , 
\end{gather}
where $\tilde{\mathcal{V}}_S(u,v,\tau)$ is the 3D delay spectrum and $P_S(u,v;\tau)$ is the power spectrum in the delay-baseline space. The subscript `S' refers to one of the Stokes parameters $I$,$Q$,$U$,$V$ or the complex polarized intensity $\mathcal{P} = Q + iU$. The 2D delay power spectrum $P_S(|b|,\tau)$ can be obtained by azimuthally averaging $P_S(u,v,\tau)$ in $uv$-plane, where $|b| = (\sqrt{u^2 + v^2})\times \lambda$ is the baseline length (in meters) and $\tau$ is the delay which corresponds to the geometric time delay between the signal arriving at two different antennas from a given direction. The delay $\tau$ can also be written as:
\begin{equation}
\tau = \dfrac{\textit{\textbf{b}}\cdot \hat{\mathbf{s}}}{c} \ ,
\end{equation}
where $\hat{\mathbf{s}}$ is the unit vector towards the direction of the incoming signal, $\theta$ is angle between zenith and $\hat{\mathbf{s}}$, and $c$ is the speed of light. For $\theta = 90^{\circ}$, $\tau = |b|/c$; this delay corresponds to the instrumental horizon.
A 2D delay spectrum scaled with proper cosmological parameters results in the 2D cosmological power spectrum, which is a widely used statistic in EoR-experiments. The 2D cosmological power spectrum can be derived from the delay spectrum as \citep{parsons2012,thyagarajan2015a}:
\begin{equation}
P(k_{\perp},k_{\parallel}) = |\tilde{\mathcal{V}}(|b|,\tau)|^2 \left( \dfrac{A_{\text{eff}}}{\lambda^2 \Delta \nu} \right) \left( \dfrac{D^2(z) \Delta D}{\Delta \nu} \right) \left( \dfrac{\lambda^2}{2k_B} \right)^2
\end{equation}
and baseline ($\textit{\textbf{b}}$) and delay ($\tau$) are related to $k_{\perp}$ and $k_{\parallel}$ wave numbers as:

\begin{equation}
k_{\perp} = \dfrac{2\pi \left( \frac{|b|}{\lambda} \right)}{D(z)} , \ \ k_{\parallel} = \dfrac{2\pi \nu_{21} H_{\text{0}} E(z)}{c(1+z)^2} \tau \ , 
\end{equation}
where $A_\text{eff}$ is the effective area of the antenna, $\lambda$ is the wavelength of the center frequency of the observation band, $\Delta \nu$ is the observation bandwidth, $D(z)$ is the transverse co-moving distance corresponding to redshift $z$, $\Delta D$ is the co-moving depth along the line of sight corresponding to $\Delta \nu$, $k_B$ is the Boltzmann constant, $\nu_{21}$ is the rest frame frequency of the 21-cm spin-flip transition of HI. $H_0$ and $E(z) \equiv \left[ \Omega_M(1+z)^3 + \Omega_k(1+z)^2 + \Omega_{\Lambda}\right]^{1/2} $ are the Hubble constant and a function of the standard cosmological parameters. We use $P(|b|,\tau)$ instead of $P(k_{\perp},k_{\parallel})$ and adhere to units of $\text{Jy}^2$ throughout our analysis. This is a suitable choice in this paper as we only address the severeness of the contamination effects, which are orders of magnitude ($\sim$ Kelvins in amplitude) higher than the expected 21-cm signal at the frequencies of interest. Typically, a delay spectrum is defined per visibility where the instrumental horizon (same as physical horizon) is fixed. In a phase tracking array, the instrumental horizon is no longer fixed and moves with respect to the zenith. Because of tracking, delays towards a particular source (fixed with respect to the phase center) in sky will vary within a certain range, depending on the orientation of baseline and location of the phase center. As a result of this, features due to that source in delay power spectrum produced using time integrated image cubes will appear to be smeared over a certain range of delays. Even in drift scan arrays, a particular source appears at a certain delay only in snapshot mode with phase center on zenith. Once the correction for earth's rotation is applied, it will appear to be smeared across several delays. 

There are some differences between the delay spectrum estimation approach in \citealt{thyagarajan2015a,thyagarajan2015b}, and the approach we used in our analysis. The former uses snapshot visibilities, that are averaged over different observing nights (same LST) to average down the incoherent part of the visibilities. These averaged visibilities are subsequently used to estimate the delay power spectra. In our case, visibilities recorded at different times during a single observation are coherently averaged during the gridding process, ultimately averaging down their incoherent (noise) part. These gridded visibilities are Fourier transformed to produce time integrated delay power spectra. 
Gridding asymptotically for large numbers of visibilities leads to the delay power spectrum of average visibilities.  Whereas, averaging of the individual visibility-based power spectra, as in \citealt{thyagarajan2015a,thyagarajan2015b}, yields the delay power spectrum of the average visibility with the power spectra of the incoherent part of the visibility (i.e. noise and scintillation noise) added to it. We opted for the delay spectrum of gridded visibilities to (i) avoid having to separately estimate each of the incoherent power spectra and (ii) reduce computational effort since diffuse foreground subtraction is computationally prohibitive if it is done at the visibility level. 

Another difference between the two approaches is that \citealt{thyagarajan2015a,thyagarajan2015b} estimate the delay power spectra directly from visibilities. Foreground subtraction in this approach will affect the power at a certain delay corresponding to the subtracted foreground source. Whereas, we determine the delay power spectra by Fourier transforming the image cubes (real-valued signal) instead of calculating it directly from the visibilities. This makes the delay transform in our case, a Hermitian transform. As an outcome of this, the resulting delay power spectrum is symmetric around $\tau = 0$ and foreground subtraction will affect the power in the same manner at positive and negative delays corresponding to the subtracted foreground source.

Estimation of $P_S(|b|,\tau)$ from  image cubes requires two additional steps. (a) The image cube is Fourier transformed along the spatial axes. The spatial axes of cosine-directions $(l,m)$ of the images are the Fourier conjugates of the baseline axes $(u,v)$. The resulting gridded visibilities for different frequencies have fixed $uv$-cell size in units of wavelength ($\lambda$) causing the physical $uv$-cell size (in meters) to vary with frequency. (2.) This physical $uv$-grid and corresponding visibilities are re-gridded on to a fixed grid with baseline length in meters such that the $uv$-coverage scales as function of frequency but the size of the physical grid remains fixed. The re-gridded visibilities are then Fourier transformed along frequency to obtain the delay spectrum. We flag several noisy sub-bands on the basis of the Stokes $V$ rms of the images to avoid any artifacts (due to RFI etc.) in the delay power spectrum. This flagging produces image cubes and hence gridded visibilities which have irregular spacing across frequency axis, and therefore a Fast Fourier Transform (FFT) cannot be used to FT across the frequency axis. Thus, we use a FFT to FT the image cube only across the spatial axes, whereas for the frequency axis, we use a Least Squares Spectral Analysis (LSSA) method (i.e. full least squares-FT matrix inversion) (see e.g. \citealt{barning1963,lomb1976,stoica2009,trott2016}). The resulting cube is then squared and azimuthally averaged (annuli width $\delta b = 19.1$ m) across the spatial domain to obtain the 2D delay power spectrum. We use the $I$, $Q$, $U$, and $V$ image cubes produced with $1000\lambda$ imaging scheme to determine the 2D delay power spectrum. 

\begin{figure*}
\centering
\includegraphics[width=0.93\textwidth]{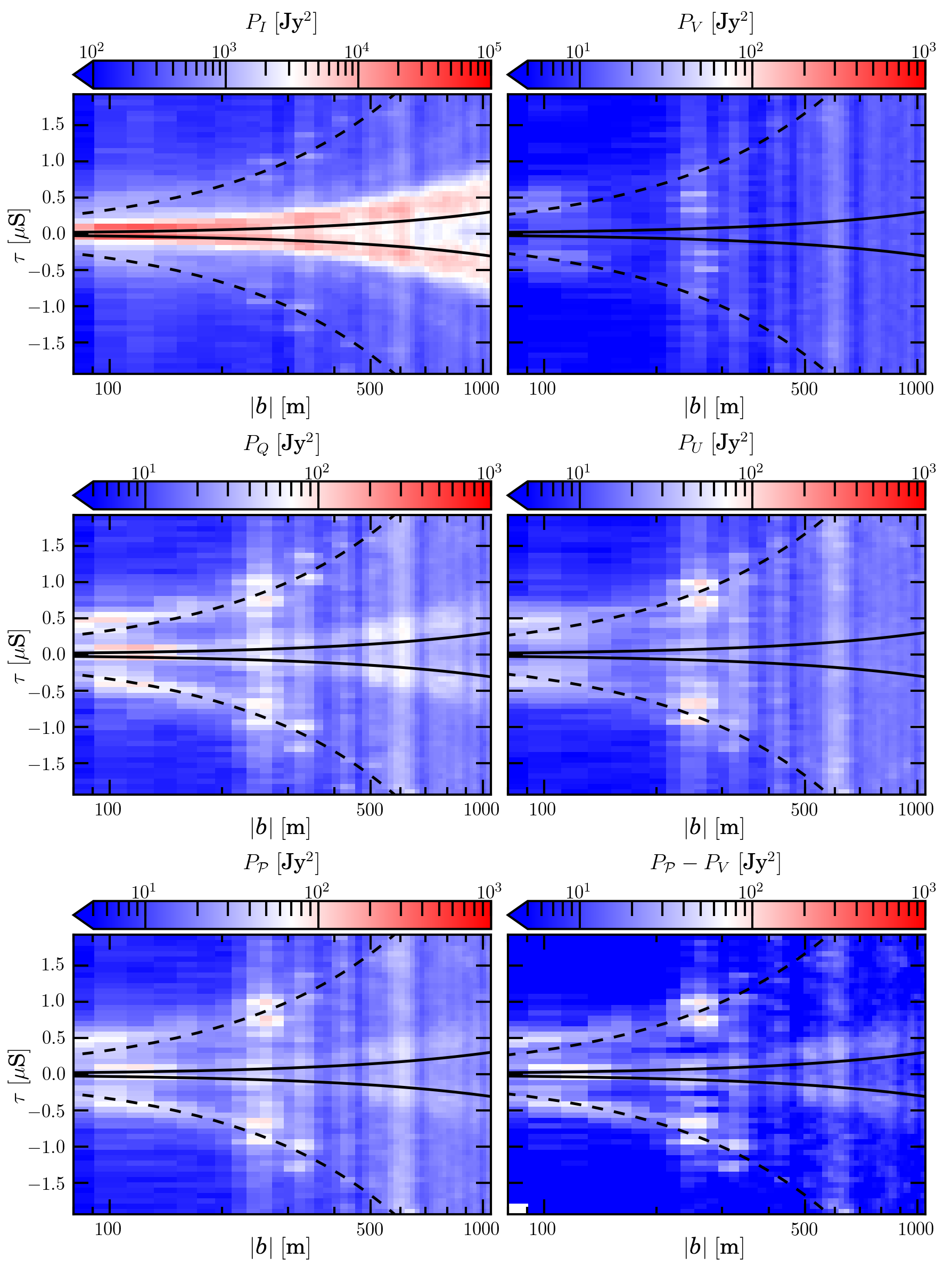}
    \caption{This figure shows the delay spectra for Stokes $I$ (top-left), $V$ (top-right), $Q$ (middle-left), $U$ (middle-right), total polarized intensity $\mathcal{P}$ (bottom-left), and the difference $P_{\mathcal{P}} - P_V$ (bottom-right). The black solid lines represent delays corresponding to FWHM of primary beam (see Table \ref{tab:obs_details}) and the dashed lines correspond to the instrumental horizon. These power spectra are computed from the image cubes produced using the $200\lambda$ DD calibrated visibilities with only {Cas\,A} subtracted. We observe a clear `pitchfork' structure in Stokes $I$,$Q$,$U$, $\mathcal{P}$ and the difference. The difference $P_{\mathcal{P}} - P_V$ show the excess of polarized emission over the Stokes $V$.}
\label{fig:IQUVPD_BBS_CasA}
\end{figure*}

\subsection{`Pitchfork' structure in polarized intensity}
We determine the delay power spectrum from Stokes $I$, $Q$, $U$, $V$ images produced after DD-calibration step using $200\lambda$ cut strategy where only {Cas\,A} is subtracted. Figure \ref{fig:IQUVPD_BBS_CasA} shows  delay power spectra for Stokes $I$, $Q$, $U$, $V$ and $\mathcal{P}$. We observe a `pitchfork' structure in Stokes $I$ power spectrum. A similar structure has been observed in MWA \citep{thyagarajan2015a,thyagarajan2015b} and PAPER \citep{kohn2016} observations. Moreover, we observe a similar `pitchfork' structure in power spectrum of Stokes $Q$, $U$ and $\mathcal{P}$. Most of this polarized emission is localized on smaller baselines ($\leq 400$ m) and around the delays corresponding to instrumental horizon, suggesting that the emission originates from far outside the primary beam and is diffuse in nature. This can either be caused by intrinsic diffuse polarized emission or instrumental polarization leakage from Stokes $I$ to $Q$ and $U$. One method to distinguish between intrinsic polarized emission and instrumental polarization leakage is to investigate the emission in Stokes $Q$, $U$ and $\mathcal{P}$ in RM-space (see Appendix A). When a polarized signal passes through an ionized medium in the presence of a net magnetic field parallel to the line of sight, the signal undergoes Faraday rotation. Due to Faraday rotation, the signal appear often at non-zero Faraday depths ($\Phi \neq 0$) in the RM-space. On the contrary, any polarization leakage due to the instrument is localized around $\Phi = 0$ because the primary beam variation has a smooth but weak dependence on the frequency. In the $\mathcal{P}$ RM-cubes, we do not see any polarized emission except at $\Phi = 0$. We expect that the polarized emission is depolarized by ionospheric Faraday rotation due to ionospheric Total Electron Content (TEC) varying as a function of time and position. The polarization angle $\chi \propto \lambda^2$. Thus, at low frequencies, ionospheric Faraday rotation becomes significant and depolarizes most of the intrinsic polarized signal. Figure \ref{fig:IQUVPD_BBS_CasA} also shows the difference $P_{\mathcal{P}} - P_{V}$ which represents the presence of excess polarized power over the power in Stokes $V$ (assumed to be the noise level). The `pitchfork' feature is also observed in the difference plot. We notice that most of the excess polarized power originates outside the primary beam and is localized around shorter baselines, i.e. $|b|< 400$ m, whereas little to no polarization power at $\tau \approx 0$ which suggests absence of  intrinsic polarized emission in the field. We confirm the absence of intrinsic polarized emission also by the noise like image cubes (not shown here) in Stokes $Q$ and $U$ (with variance exceeding Stokes $V$ though). We also observe a faint structure in Stokes $V$ which appear to correlate with the `pitchfork' structure in Stokes $Q$ and $U$ on certain baselines. Because there is negligible emission (circularly polarized component) in Stokes $V$, it is expected to have a flat power spectrum. Presence of any structure in Stokes $V$ is another indication of instrumental polarization leakage. 

\subsubsection{Effect of calibration cut}
\begin{figure*}
\centering 
\includegraphics[width=0.9\textwidth]{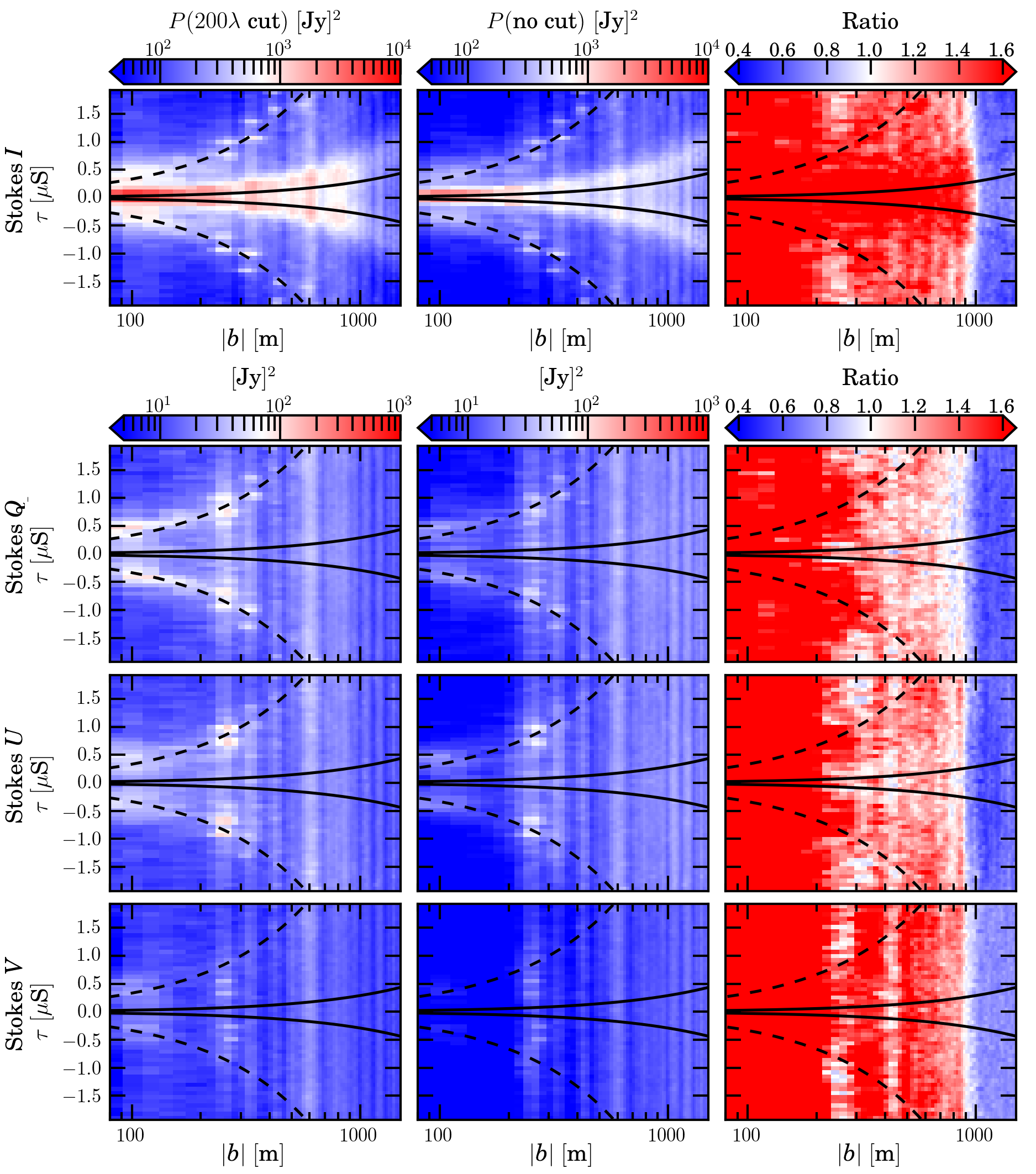}
    \caption{Stokes $I$, $Q$, $U$ and $V$ delay spectra for two different DD calibration schemes and their ratio. Leftmost column corresponds to DD calibration scheme using $\geq200 \lambda$ baselines, middle column corresponds to DD calibration scheme using all baselines and rightmost column corresponds to the ratio $\frac{P(200\lambda \text{ cut})}{P(\text{no cut}}$. Different rows correspond to different Stokes parameters. Top row corresponds to Stokes $I$, second row from top corresponds to Stokes $Q$, third row from top corresponds to Stokes $U$ and bottom row corresponds to Stokes $V$.}
\label{fig:IQUV_calib_compare}
\end{figure*}

To quantify the impact of DD calibration on the `pitchfork', we used the visibilities after DD calibration step where {Cas\,A} and the in-field model has been subtracted. We compare the power spectra of visibilities from the two calibration strategies (Table \ref{tab:cal_params}). Figure \ref{fig:IQUV_calib_compare} shows Stokes $I$, $Q$, $U$ and $V$ delay power spectra from the two calibration strategies and their ratio ($P(200\lambda \text{ cut})/P(\text{no cut})$). We notice that the power on/around the `pitchfork' is suppressed significantly when all the baselines are used in calibration. This might happen because the unmodeled diffuse emission is absorbed in the gain solutions, hence lowering the power on smaller baselines (\citealt{patil2016}, Sardarabadi et al. in prep).

We observe a discontinuity in the ratio at $|b| \sim 900$ m, the ratio drops for $|b| > 900$ m and continues to drop till $|b| \sim 1000$ m and becomes almost constant for $|b| > 1000$ m. The $200\lambda$ baseline cut for different frequency sub-bands lies in baseline range $900 \text{ m} < |b| < 1000 \text{ m}$. Therefore, this discontinuity around $900 \text{ m} < |b| < 1000 \text{ m}$ corresponds to the location of baseline cut and is similar to the one in the ratio of differential power spectrum (figure \ref{fig:dfferentialspectra}, right panel). This trend is observed for all the Stokes parameters. We observe that the excess power on excluded short baselines is $\gtrsim 2$ times the power on baselines included in the calibration step. A similar reasoning, as in section 3.2, can be applied in this case as well; that power on baselines $|b|<200\lambda$ is enhanced because of the errors in the gain solutions (obtained solely from the longer baselines) applied to the data and to the sky-model. The source of these errors is not yet well understood, but we suspect several causes such as incomplete calibration models, ionospheric effects and imperfect calibration \citep{patil2016,barry2016}. 

\begin{figure*}
\centering
\includegraphics[width=0.9\textwidth]{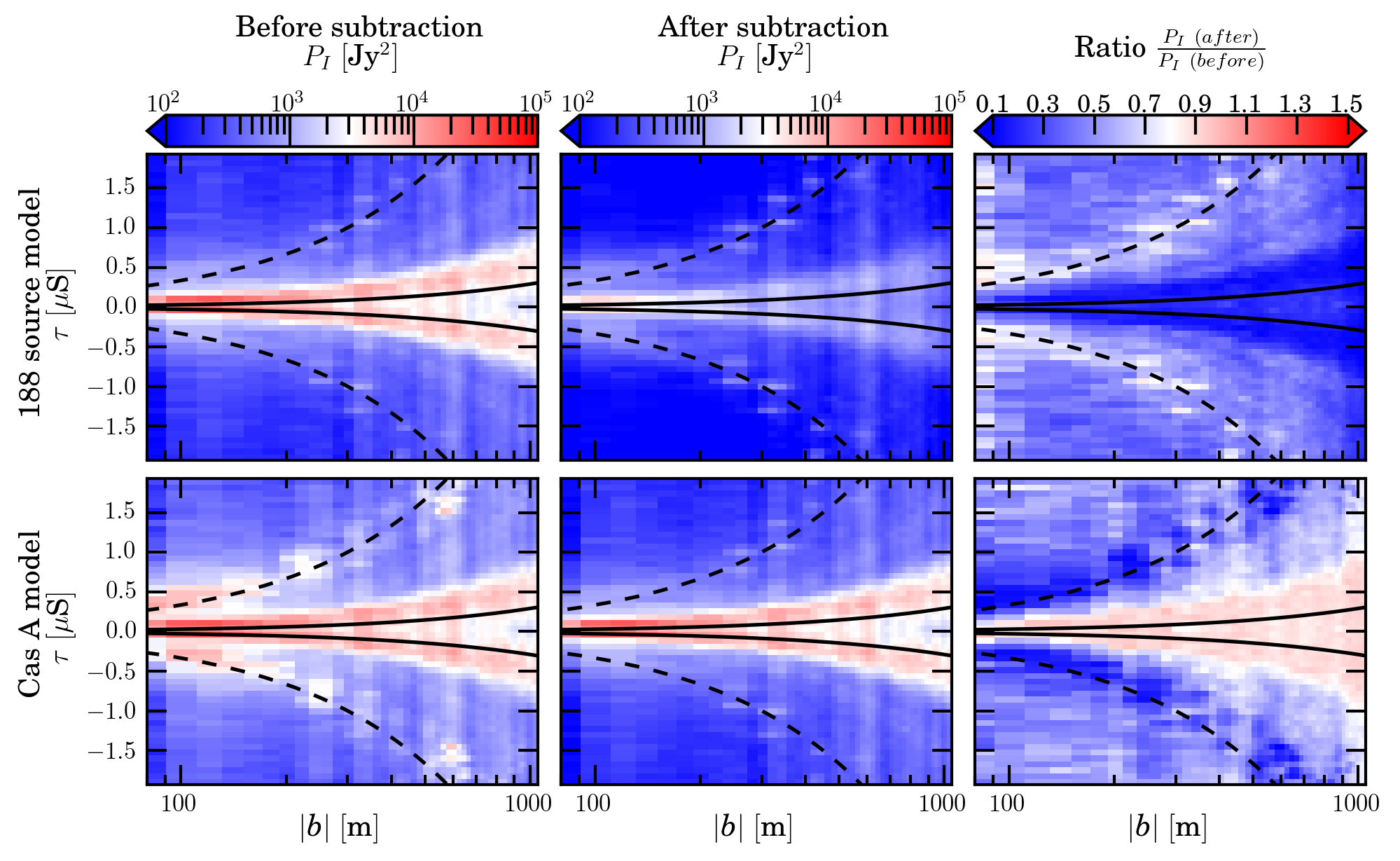}
    \caption{Stokes $I$ delay power spectra before (left column) and after (center column) model subtraction and their ratio $(P_I(\text{after})/P_I(\text{before}))$ (right column). The top row shows $P_{I}$ before and after in-field model subtraction and their ratio. The bottom row shows $P_{I}$ before and after {Cas\,A} model subtraction and their ratio.}
\label{fig:I_compare_modelsub}
\end{figure*}

\begin{figure*}
\centering
\includegraphics[width=0.9\textwidth]{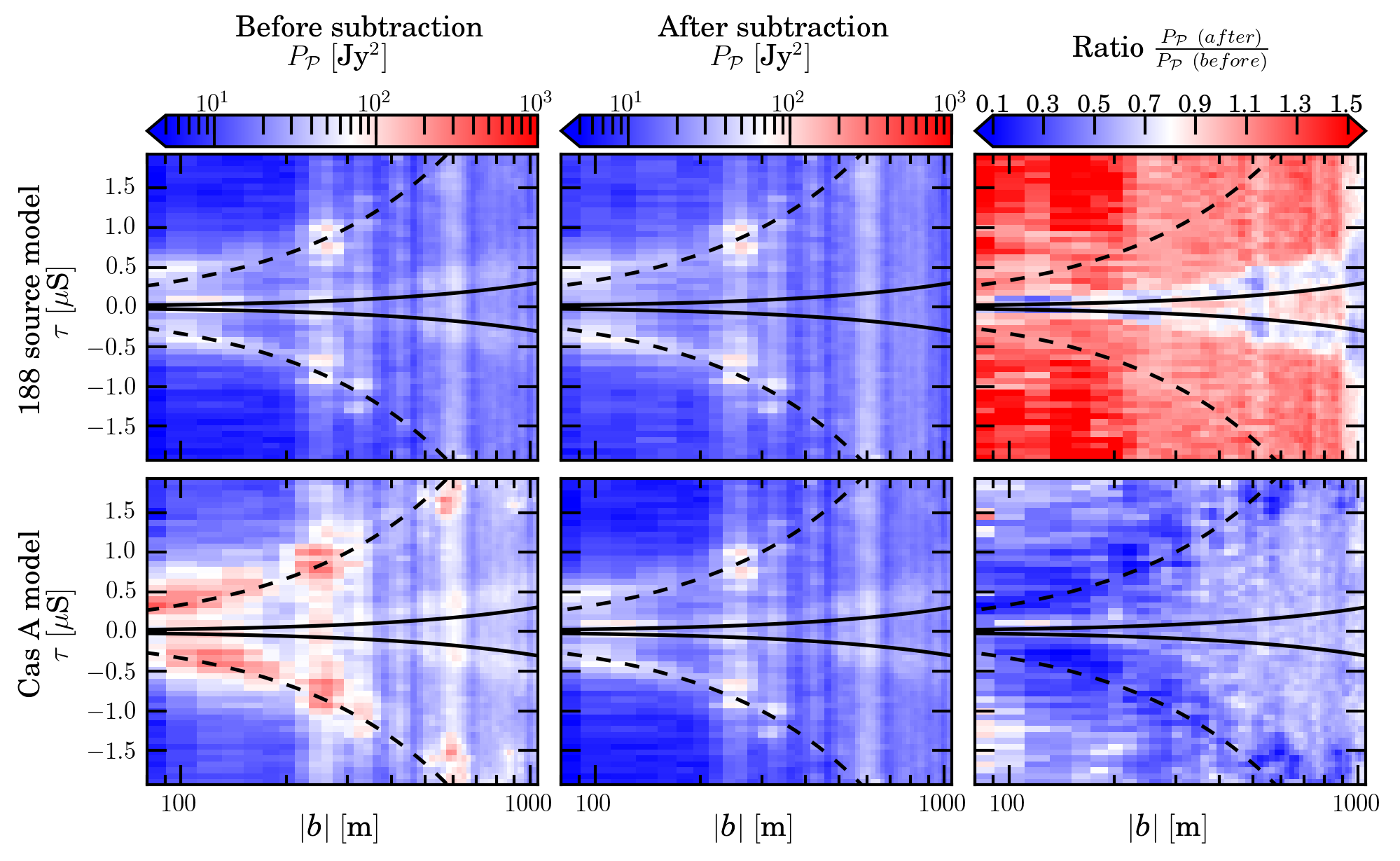}
    \caption{Polarized intensity $\mathcal{P}$ delay power spectra before (left column) and after (center column) model subtraction and their ratio (right column). The top row shows $P_{\mathcal{P}}$ before and after in-field model subtraction and their ratio. The bottom row shows $P_{\mathcal{P}}$ before and after {Cas\,A} model subtraction and their ratio.}
\label{fig:P_compare_modelsub}
\end{figure*}

\subsubsection{Effect of source subtraction} 
In this section, we discuss the effect of subtraction of sources on the `pitchfork' structure. We quantify this effect for two cases. In first case, the in-field sky-model (sources within the primary beam) is subtracted from DI-calibrated visibilities ($250 \lambda$ cut) using the DD calibration ($200 \lambda$ cut). Note that {Cas\,A} model is already subtracted before performing the in-field model subtraction. We compare the delay power spectrum of Stokes $I$ ($P_I$) and $\mathcal{P}$ ($P_{\mathcal{P}}$) calculated using the image cubes before and after subtracting the model. Top row of figure \ref{fig:I_compare_modelsub} shows $P_I$ before and after in-field model subtraction and the ratio $P_I(\text{after})/P_I(\text{before})$. We observe that subtracting the sources largely within the primary beam significantly reduces the power in Stokes $I$ within the primary beam going up till the horizon as well as above the horizon. This effect is expected as a consequence of foreground subtraction. However, the ratio on/around the `pitchfork' remains $\sim 0.8-0.9$, suggesting that the subtraction of sources within primary beam does not affect the `pitchfork'. We observe a similar effect in comparison of $P_{\mathcal{P}}$ before and after in-field model subtraction. Figure \ref{fig:P_compare_modelsub} (top row) shows $P_{\mathcal{P}}$ before and after subtracting the in-field model. We observe $\sim 30\%$ decrease in polarized power within the primary beam primarily due to subtraction of sources away from phase center, but it does not affect the power beyond the primary beam. However, we observe an increase in ratio ($P_{\mathcal{P}}(\text{after})/P_{\mathcal{P}}(\text{before}) \gtrsim 1.0$) beyond horizon on baselines $\leq 200$ m. We suspect that this increase in power is due to errors on gain solutions obtained in DD calibration step, which mainly affect the shorter baselines excluded from calibration. This effect is not visible in Stokes $I$, as the subtracted power is much larger than the increase in power introduced due to these gain errors. Whereas in $\mathcal{P}$, power on excluded baselines is comparable to the increase in power introduced due to errors on gain solutions and it becomes prominent in the ratio. Besides this, we do not observe any significant difference in $P_{\mathcal{P}}$ due to the subtraction of sources within primary beam.   

\begin{figure*}
\centering
\includegraphics[width=\textwidth]{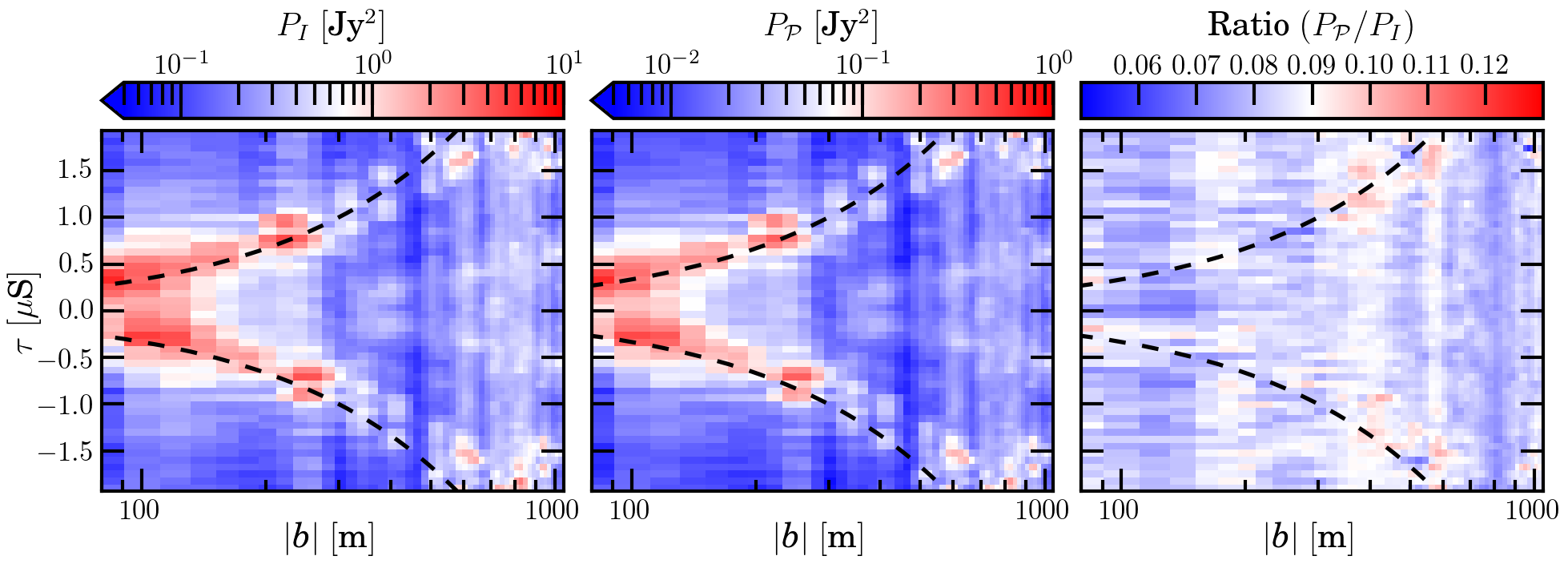}
    \caption{Stokes $I$ (left panel) and Polarized intensity $\mathcal{P}$ (middle panel) delay power spectrum determined using simulated visibilities (see section 4.1.3) and the ratio $P_{\mathcal{P}}/P_I$ (right panel). Note that the amplitude scales are apparent and are only meant to compare the power between $I$ and $\mathcal{P}$.}
\label{fig:SimulateCasA}
\end{figure*}

In second case, we compare delay power spectra for Stokes $I$ ($P_I$) and $\mathcal{P}$ ($P_{\mathcal{P}}$) before and after subtracting {Cas\,A} (using DD calibration) which lies outside the primary beam. In this case, we do not subtract the in-field model. In our observation, {Cas\,A} is above horizon during the whole period of observation and is $\gtrsim 40^{\circ}$ away from the zenith ($\sim 66^{\circ}$ away from 3C196; $\sim 31^{\circ}$ away from NCP). Figure \ref{fig:I_compare_modelsub} (bottom row) and figure \ref{fig:P_compare_modelsub} (bottom row) show $P_I$ and $P_{\mathcal{P}}$ respectively before and after {Cas\,A} subtraction and the ratio ($P_I(\text{after})/P_I(\text{before})$). We observe a factor of $\sim10$ decrease in power on the `pitchfork' in both Stokes $I$ and $\mathcal{P}$ after {Cas\,A} subtraction. From this comparison, it is clear that subtraction of {Cas\,A} has a significant impact on the power in Stokes $I$ and polarized intensity $\mathcal{P}$ on/around the `pitchfork' but also on the modes within and beyond the horizon ($\sim 50\%$ decrease). Since {Cas\,A} is extremely bright at low frequencies ($\sim21$ kJy intrinsic flux at 81 MHz; \citealt{baars1977}), its effects can be detected in LBA images even when it is tens of degrees away from the phase center. LOFAR-LBA has a polarized response for angles away from zenith. For zenith angles $\approx 60^{\circ}$, $\mathcal{P}/I \approx 0.3$ (see e.g. \citealt{bregman2012}), causing significant fraction of the total power ($\sim10 \%$) leak to polarized power due to the instrument. This leakage occurs from $\mathcal{P}$ to $I$ as well. Note that most of the leaked power is on the small baselines ($<600$ m), which is probably due to the large extent of {Cas\,A} caused by ionospheric diffraction (discussed later). The leakage is reduced substantially when {Cas\,A} is subtracted using a model via DD calibration. Note that residuals after subtracting {Cas\,A} still correlate quite strongly with the power before {Cas\,A} subtraction, suggesting imperfect subtraction in DD calibration or the structure of {Cas\,A} which is harder to model. In summary, the primary cause of the `pitchfork' structure in $\mathcal{P}$ is {Cas\,A} outside the primary beam leaking to $\mathcal{P}$ from Stokes $I$ because of the instrumental beam polarization. Although other sources which are spread over many directions (and delays) will also leak in to $\mathcal{P}$ as shown in \cite{asad2016,asad2017}, they are unlikely to cause strong leakage. A single source as bright as {Cas\,A} however, is clearly dominant in the power spectra. 

\subsubsection{Comparison with the simulations}
To gain further insight on the `pitchfork' structure, we simulate visibilities observed by LOFAR-LBA using a Stokes $I$ only model of {Cas\,A}, with the phase center at 3C196. We use \texttt{NDPPP} \footnote{\url{http://www.lofar.org/operations/doku.php?id=public:user_software:ndppp}} to predict the $XX$, $XY$, $YX$, $YY$ antenna correlations using the exact LOFAR-LBA station configuration for 4 hours of synthesis. We chose the time and frequency resolution of the correlations to be 5 seconds and 183.1 kHz to save computation time. We include the LOFAR-LBA primary beam (a recent addition to LOFAR data processing pipeline (\texttt{NDPPP}), see footnote 12) in the prediction step in order to predict instrumental polarization leakage. We then image the predicted visibilities using \texttt{WSClean} using the $1000\lambda$ imaging scheme and determine the delay power spectrum for Stokes $I$ and total polarized intensity $\mathcal{P}$. Figure \ref{fig:SimulateCasA} shows Stokes $I$ and $\mathcal{P}$ delay power spectrum and the ratio $P_{\mathcal{P}}/P_I$. We observe a clear `pitchfork' structure in $P_I$ and this structure appears solely due to {Cas\,A}. The structure looks nearly identical to that observed in figures \ref{fig:I_compare_modelsub} and \ref{fig:P_compare_modelsub}. If we compare $P_I$ with $P_{\mathcal{P}}$, the structure in $P_{\mathcal{P}}$ looks exactly like that in $P_I$, but scaled down in power. The ratio $\frac{P_{\mathcal{P}}}{P_I} \approx 0.09$, which is $\sim 0.3$ in amplitude. The effect of a constant ratio between Stokes $I$ and ${\mathcal{P}}$ due to polarization leakage was also predicted by \cite{asad2017} for LOFAR-HBA observations. This simulation clearly shows that the `pitchfork' structure in $\mathcal{P}$ is indeed an artifact arising from {Cas\,A} due to instrumental polarization leakage from Stokes $I$ to $\mathcal{P}$.

We also simulate the visibilities using a {Cyg\,A} only model to quantify the polarization leakage due to {Cyg\,A}. We used the VLSS model of {Cyg\,A} ($\sim20$ kJy at 74 MHz \citep{kassim2007}) with the same simulation setup as for {Cas\,A} to predict the antenna correlations. The Stokes $I$ power spectrum (not shown here) calculated using the simulated visibilities for {Cyg\,A} shows $\sim$6-7 orders of magnitude lower power on the `pitchfork' compared to the power due to {Cas\,A}. Although the beam model used in simulations is only approximately correct (inaccuracy of $\sim$ few percent) in the direction of {Cyg\,A} (lower elevation angles), contribution due to {Cyg\,A} is negligible and can be ignored for any practical purpose in observations with LOFAR-LBA centered on 3C196, at the current level of accuracy.

When the model of {Cas\,A} is subtracted from the visibilities during the DD calibration step, the `pitchfork' structure due to {Cas\,A} should in principle (if the model is accurate) disappear. However, we still observe some residual power on/around the pitchfork. The residual power on small baselines ($\leq 400$ m) is $\sim 10\%$ of the power before {Cas\,A} subtraction. These residuals can be caused by other factors such as unmodeled sources, diffuse emission, an inaccurate {Cas\,A} model, imperfect calibration and ionospheric effects. For example, {Cas\,A} is 3 arcmin in extent, which should be resolved only on the baselines $>800\lambda$. {Cas\,A} should therefore appear approximately as a compact source on baselines $\leq 100\lambda$. Thus, inaccuracy in the {Cas\,A} model should not cause such significant residuals on these baselines. Ionospheric turbulence, on the other hand, can cause {Cas\,A} to scintillate significantly and visibilities to decorrelate within the DD calibration solution interval. On baselines $<5$ km, the ionosphere decorrelates on time scales of less than a minute, which is shorter than the solution interval in the DD calibration (5 minutes). Therefore, this `scintillation noise' (see e.g. \citealt{vedantham2015,vedantham2016}) might lead to imperfect calibration causing residual flux. We discuss this effect in the next section.

\begin{figure}
\centering
\includegraphics[width=\columnwidth]{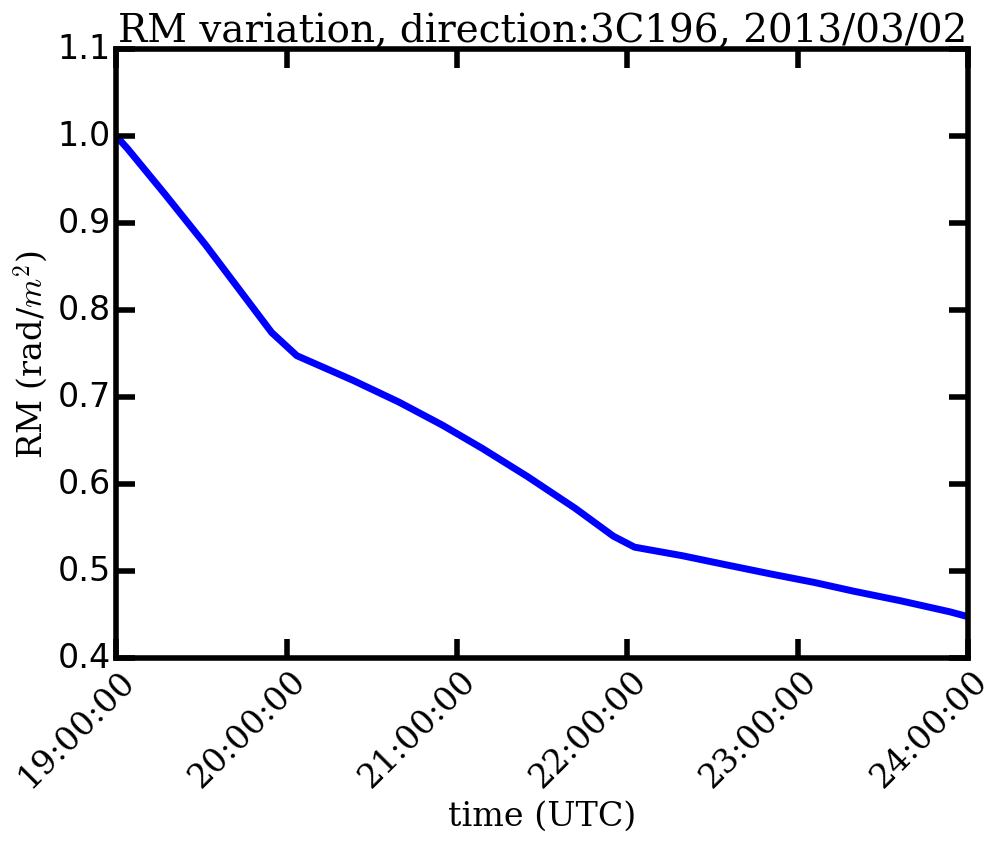}
    \caption{Ionospheric RM variation in direction of 3C196 as a function of time on March 2, 2013. This variation is calculated using \texttt{RMextract} developed by Maaijke Mevius. \texttt{RMextract} uses the GPS data to extract RM in particular direction.}
\label{fig:RMvaration}
\end{figure}

\section{Ionospheric Scintillation}
In previous analysis, we observed that the $\mathcal{P}$ delay power spectrum has more power concentrated on the `pitchfork' on smaller baselines ($<400$ m). This feature is present in delay power spectra for both calibration strategies, and is associated with polarization leakage from {Cas\,A}. There is residual flux in $\mathcal{P}$ delay power spectrum even after subtraction of {Cas\,A}. Given the low-frequency and large angle away from zenith (i.e. large vTEC), we expect {Cas\,A} to be strongly affected by the ionosphere. The ionospheric turbulence is usually carried along with the bulk motion of ionospheric plasma, which has typical speeds between $100 \  \text{to} \ 500 \ \text{km/h}$. Turbulent plasma in the ionosphere introduces time, frequency and position dependent phase shifts to the propagating wave. Under the phase-screen approximation, the phase shift $\phi$ introduced due to the wave propagation through ionospheric plasma is 
\begin{equation}
\phi = \int \dfrac{2\pi \eta(z)}{\lambda}dz  \ ,
\end{equation}
where $z$ is the distance along the direction of propagating wave. $\eta$ (refractive index of non-magnetized plasma) is given by
\begin{equation}
\eta = \sqrt{1 - \dfrac{\nu_p^2}{\nu^2}} \approx 1 - \dfrac{1}{2}\dfrac{\nu_p^2}{\nu^2} \ \text{for} \ \nu_p << \nu \ ,
\end{equation}
where $\nu_p$ is the plasma frequency (order of few MHz) and $\nu$ is the frequency of the propagating wave. By combining equation 15 and 16, $\phi$ can be written as  
\begin{equation}
\phi = \int \dfrac{2\pi \eta(z)}{\lambda}dz = \int \dfrac{2\pi \nu}{c}dz - \dfrac{1}{2} \int \dfrac{2\pi \nu_p^2}{c\nu}dz \ . 
\end{equation}

\begin{figure}
\centering
\includegraphics[width=\columnwidth]{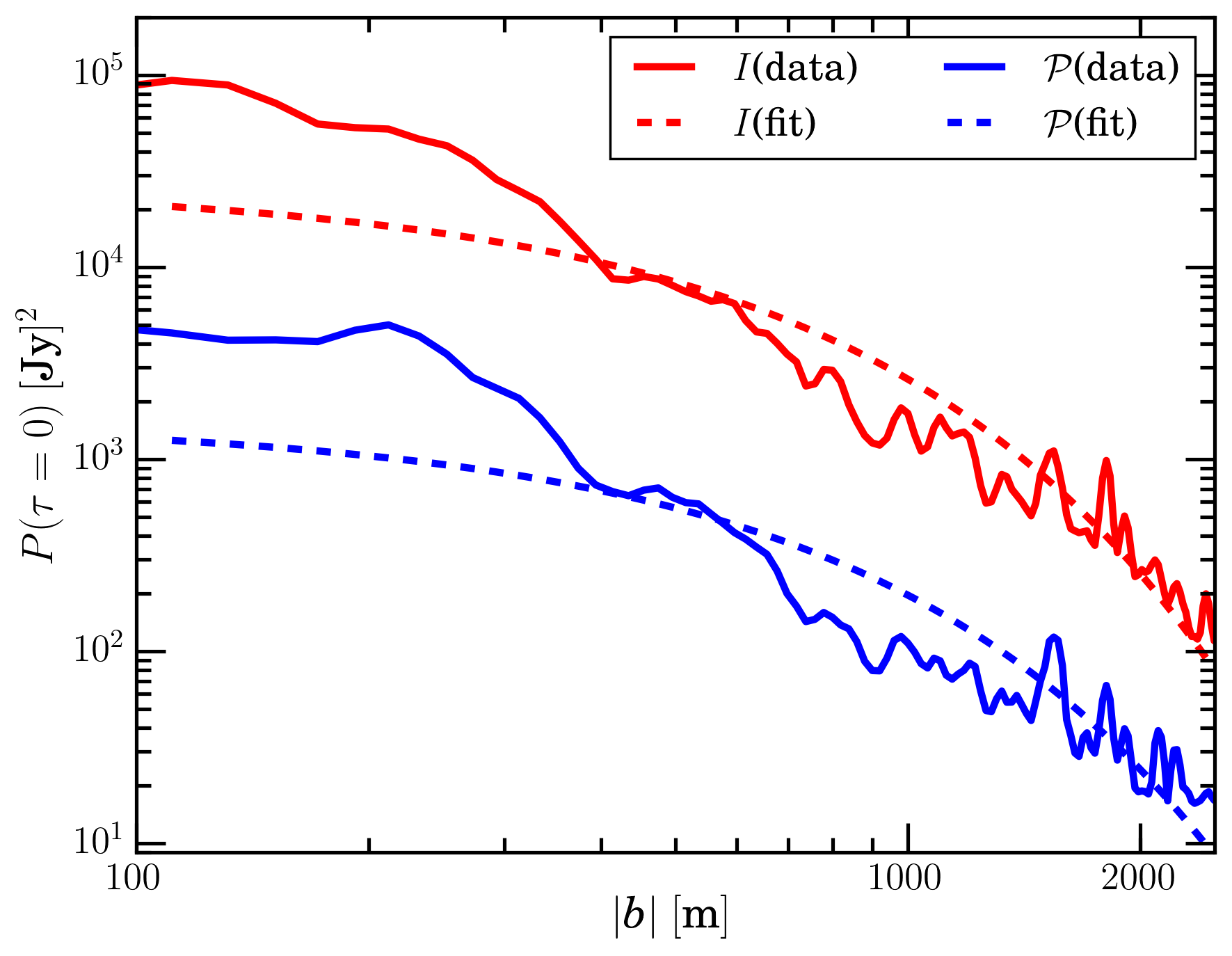}
    \caption{This figure shows $\tau = 0$ slice of Stokes $I$ (solid-red curve) and $\mathcal{P}$ (solid-blue curve) delay power spectra determined using the DI calibrated visibilities ($250\lambda$ cut) which are phase rotated towards {Cas\,A}. The dashed curves show power spectrum for pure Kolmogorov turbulence (equation 22) with best fitting values listed in Table \ref{tab:kolfit_params} for Stokes $I$ (red) and $\mathcal{P}$ (blue).}
\label{fig:zero_delay_PS_fit}
\end{figure}

The first term in equation 17 is a geometric delay term which is generally absorbed in the interferometer measurement equation. The second term in equation 17 is inversely proportional to the frequency of the propagating wave ($\nu$): 
\begin{equation}
\phi(\nu) \propto \dfrac{\nu_p^2}{\nu}, \ \text{where} \ \nu_p = \dfrac{1}{2\pi} \sqrt{\dfrac{n_e q_e^2}{m_e \epsilon_{\circ}}} \ ,
\end{equation}
$n_e$ is plasma density, $q_e$ is the electron charge, $m_e$ is the mass of electron and $\epsilon_{\circ}$ is the permittivity of free space. The spatial variations in $n_e$ can be described by Kolmogorov-type turbulence \citep{rufenach1972,singleton1974,koopmans2010,vedantham2015}. The power spectrum for Kolmogorov-type turbulence is represented by a $-11/3$ index power law. Since $\phi \propto n_e$ (follows from equation 18), the phase fluctuations are also described by a Gaussian random field with power spectrum (assuming isotropy) given by	
\begin{equation}
|\tilde{\phi}(k)|^2 \propto  k^{-11/3}, \ k_o < k < k_i
\end{equation}
where $k$ is the length of the spatial wavenumber vector $\textbf{k}$, $k_o$ is the wavenumber corresponding to the outer scale or the energy injection scale, and $k_i$ corresponds to the inner scale or energy dissipation scale. If the visibility of a source in absence of ionospheric effects is given by $\mathcal{V}_S(\textbf{\textit{b}})$, then the expectation value of visibilities corrupted by the ionospheric phase fluctuations (assuming that the calibration solution interval significantly exceeds the time scale on which the phases fluctuate) is given by (see e.g. \citealt{vedantham2015,vedantham2016}):
\begin{equation}
\langle \mathcal{V}_C(\textbf{\textit{b}}) \rangle = \mathcal{V}_S(\textbf{\textit{b}}) \ \text{exp} \left( - \dfrac{1}{2}\mathcal{D}(b) \right) 
\end{equation}
where $\mathcal{V}_C(\textbf{\textit{b}})$ are the time-averaged visibilities. $\mathcal{D}(b)$ is the phase structure function and is defined as:
\begin{equation}
\mathcal{D}(b) = \left( \dfrac{b}{r_{\text{diff}}} \right)^{5/3} ,
\end{equation}
where $r_{\text{diff}}$ is the diffractive scale. The power spectrum of the visibilities corrupted by the ionosphere is given by:
\begin{equation}
P_C(b)  = |\mathcal{V}_S(\textbf{\textit{b}})|^2 \ \text{e}^{-\mathcal{D}(b)} = P_{S} \text{ exp} \left[ - \left( \dfrac{b}{r_{\text{diff}}} \right)^{5/3} \right] 
\end{equation}
The power spectrum of an unresolved source as a function of baselines is constant in absence of ionospheric effects, whereas if the source is affected by the ionospheric phase fluctuations, it will take the form of $P_C(b)$. To determine $P(|b|,\tau=0)$, we selected DI calibrated visibilities with $250\lambda$ cut strategy. We phase rotate these visibilities towards {Cas\,A} and image them with the $1000\lambda$ scheme. We used the resulting image cubes to obtain $P_I(|b|,\tau)$ and $P_{\mathcal{P}}(|b|,\tau)$. We then choose the $\tau = 0$ slice from each $P_I(|b|,\tau)$ and $P_{\mathcal{P}}(|b|,\tau)$, which are expected to be dominated by the power due to {Cas\,A}, and fit them with $P_C(b)$ in equation 22 using $P_S$ and $r_{\text{diff}}$ as free parameters. We use $100 \ \text{m} \ \leq |b| \leq \ 2500 \ \text{m} $ baselines for fitting. Because {Cas\,A} exhibits a 3 arcmin structure and is only resolved on baselines $>800 \lambda$ ($\sim 4$ km at 60 MHz), means that the intrinsic power spectrum for {Cas\,A} is flat for selected baselines. Figure \ref{fig:zero_delay_PS_fit} shows the $P_{I}(|b|,\tau=0)$ and $P_{\mathcal{P}}(|b|,\tau=0)$ slices fitted with equation 22.

We can see that equation 22 fits $P_I(|b|,0)$ and $P_{\mathcal{P}}(|b|,0)$ for over three orders of magnitude in power. The best-fitting values for $P_S$ and $r_{\text{diff}}$ for both power spectra are listed in Table \ref{tab:kolfit_params}. We find a diffractive scale $r_{\text{diff}}$ towards {Cas\,A} of order $\sim 430$ m for $P_{I}$ and $\sim 480$ m $P_{\mathcal{P}}$. Estimated $r_{\text{diff}}$ values for $P_{I}$ and $P_{\mathcal{P}}$ agree with each other within $10\%$ error. Typical values of $r_{\text{diff}}$ at zenith vary between 3 km to 20 km at 150 MHz and scale with frequency as $r_{\text{diff}} \propto \nu^{6/5}$ \citep{mevius2016,vedantham2016} and varies between 1 km to 10 km for zenith at 60 MHz. Therefore, the diffractive scales we have measured are the smallest scales ever measured at $\sim 60$ MHz. $P_{\mathcal{P}}$ has the same $r_{\text{diff}}$ (within the errors) as $P_{I}$ but is scaled down by one order of magnitude in power. This is additional evidence of instrumental polarization leakage from Stokes $I$ to $\mathcal{P}$. The ratio of the power $P_{\mathcal{P}}/P_{I} \sim 0.1$ is approximately same as the estimate of the polarization leakage for {Cas\,A} obtained in simulation results shown in section 4.1.3. We also observe that $P_{I}$ and $P_{\mathcal{P}}$ deviate from the fit for $|b| < 400$ m, which corresponds to the Fresnel scale ($r_F \sim 400$ m at 60 MHz, see e.g \citealt{vedantham2015,vedantham2016} for more details). Baselines below Fresnel scale are dominated by amplitude scintillation whereas the baselines above Fresnel scale are dominated by phase scintillation producing a better fit on $|b| > 400$ m compared to $|b| < 400$ m.

\begin{table}
	\centering
	\caption{Best-fitting Parameters}
	\label{tab:kolfit_params}
	\begin{tabular}{lll} 
		\hline 				
		\textbf{Fit parameters} & $P_{I}(|b|,\tau=0)$ slice & $P_{\mathcal{P}}(|b|,\tau=0)$ slice \\
		\hline		
		$P_S$ ($Jy^2$)            & $26775.1\pm 3010.6$ & $1580.5 \pm 209.1$ \\
		$r_{\text{diff}}$ (in $\lambda$) & $429.8\pm 13.3$	& $479.3 \pm 19.3$ \\	
		\hline
	\end{tabular}
\end{table}

\begin{figure*}
\centering
\includegraphics[width=\textwidth]{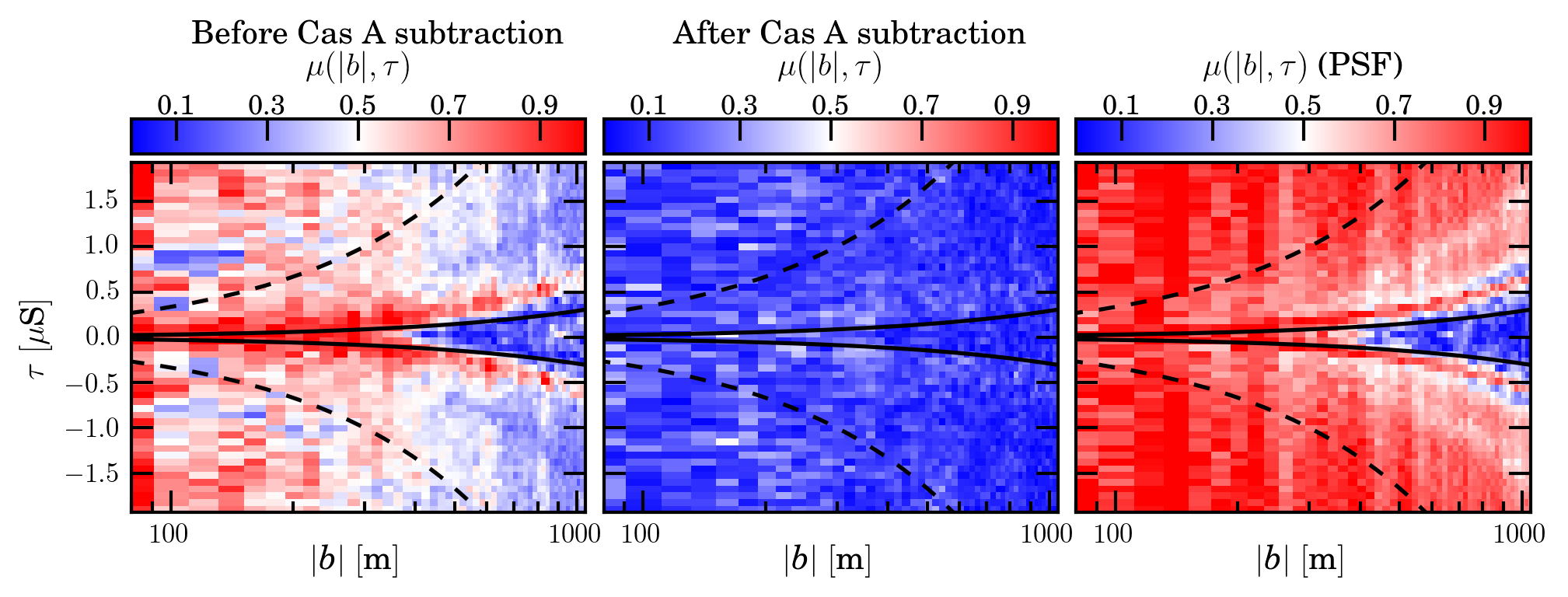}
    \caption{Stokes $I$ cross coherence ($\mu(|\mathbf{b}|,\tau)$) in delay baseline space before (left panel) and after (middle panel) subtracting {Cas\,A}. Right panel shows the PSF cross coherence $\mu_{PSF}$.}
    \label{fig:delay_crosscoherence}
\end{figure*}

For time scales of 5 minutes which correspond to the solution interval of DD calibration, ionospheric scintillation is expected to decorrelate on baselines $<1000\lambda$. We calculate the cross coherence ($\mu(|b|,\tau)$) to quantify the decorrelation of this scintillation. To determine $\mu(|b|,\tau)$, we select visibilities before {Cas\,A} subtraction, and visibilities after {Cas\,A} subtraction. We arrange each visibility set in subsets of 5 minutes duration, such that each subset corresponds to a different DD calibration solution. Next, we divide the visibility set in two non-overlapping consecutive subsets such that one subset consists of visibilities ($\mathcal{V}_{\text{odd}}$) corresponding to \texttt{odd} numbered calibration solutions and the other subset consists of visibilities ($\mathcal{V}_{\text{even}}$) corresponding to \texttt{even} numbered calibration solutions. The resulting visibility subsets, $\mathcal{V}_{\text{odd}}$ and $\mathcal{V}_{\text{even}}$ are interleaved in time. We phase rotate the visibilities towards {Cas\,A} and image them to get the corresponding image cubes $I_{\text{odd}}$ and $I_{\text{even}}$. We calculate the cross coherence ($\mu(|\mathbf{b}|,\tau)$) i.e. the normalized cross power spectrum as:
\begin{equation}
\mu(|b|,\tau) = \dfrac{|\mathcal{V}_{\text{even}} \mathcal{V}^*_{\text{odd}}|}{\sqrt{|\mathcal{V}_{\text{even}}|^2 |\mathcal{V}_{\text{odd}}|^2}} = \dfrac{|\tilde{I}_{\text{even}} \tilde{I}^*_{\text{odd}}|}{\sqrt{|\tilde{I}_{\text{even}}|^2 |\tilde{I}_{\text{odd}}|^2}}.
\end{equation}
To determine $\mu(|b|,\tau)$, we calculate the 3D power spectra  $|\tilde{I}_{\text{even}}|^2$ and $|\tilde{I}_{\text{odd}}|^2$ and the cross-power spectrum $|\tilde{I}_{\text{even}} \tilde{I}^*_{\text{odd}}|$. We perform azimuthal averaging to obtain the corresponding delay power spectra. Finally, we use these delay power spectra to calculate $\mu(|b|,\tau)$ in delay-baseline space. Figure \ref{fig:delay_crosscoherence} shows the cross coherence in the direction of {Cas\,A} before and after subtracting {Cas\,A}. Left panel of figure \ref{fig:delay_crosscoherence} shows the cross coherence between $I_{\text{even}}$ and $I_{\text{odd}}$ in direction of {Cas\,A}. We observe that $\mu(|b|,\tau)\sim 0.8-1.0$ for $|b| \lesssim 400$ m and drops afterwards. The middle panel of figure \ref{fig:delay_crosscoherence} shows $\mu(|b|,\tau)$ after the subtraction of {Cas\,A} with its DD gain solutions. We notice that effectively all correlation disappears, suggesting that most of the {Cas\,A} residuals seen in figure \ref{fig:P_compare_modelsub} are incoherent over 5 min intervals as expected for ionospheric scintillation noise.

Figure \ref{fig:delay_crosscoherence} (right panel) shows $\mu_{PSF}(|b|,\tau)$ in delay baseline space. Note that the incoherent structure inside primary beam delay line on $|b| > 400$ m in $\mu(|b|,\tau)$, before {Cas\,A} subtraction, correlates with the structure at the same location in $\mu_{PSF}(|b|,\tau)$. We can attribute this structure to the migration of baselines from one $uv$-cell to another in 5 minute timescale. In the $uv$-plane, a typical baseline vector $\textbf{b}(u,v)$ with small baseline length will traverse a smaller distance in a given time compared to a baseline vector with larger baseline length. This migration of baseline vector across the $uv$-plane mixes with the frequency dependence of the baseline vector to produce this incoherence effect in delay-baseline space. This effect is purely a $uv$-plane sampling effect and appears in cross coherence between $PSF_{\text{even}}$ and $PSF_{\text{odd}}$. 

\section{Conclusions and summary}
The LOFAR-EoR project aims at statistical detection of HI signal from redshifts $z = $ 7-12 and to measure the 21-cm power spectrum as a function of redshift \citep{patil2017}. LOFAR also operates at frequencies corresponding to the Cosmic Dawn (CD), making it in principle possible to measure or set limits on the CD power spectrum using the LOFAR-LBA system. Several contamination effects such as foreground contamination, instrumental polarization, ionospheric effects, calibration effects etc. make the detection of redshifted 21 cm emission from neutral hydrogen at high redshifts an extremely challenging task. These contamination effects are orders of magnitude stronger than the expected signal in terms of the brightness temperature. Therefore, understanding the nature of these contaminants and how they corrupt the 21 cm power spectrum becomes a crucial step in the calibration and signal extraction process. In this paper, we use several techniques such as the differential power spectrum, delay power spectrum and cross coherence to study various contamination effects in LOFAR-LBA data at low frequencies (56-70 MHz). The main results of the paper are summarized below.

\begin{enumerate}
\item We find that the excess power in the differential power spectrum of Stokes $I$ is $\sim10$ times larger than that of Stokes $V$. A similar behavior has been observed in HBA observations but it is by far not as severe as we observe in our analysis. This ratio is almost flat and does not change between the two calibration strategies with or without a baseline cut (i.e. using $|b|\geq 200 \lambda$ or using all baselines in calibration), even though the power spectra themselves change. The reasons for this excess power might be incomplete sky-model, ionospheric effects and/or imperfect calibration.

\item Introducing a baseline cut in calibration decreases the power on baselines outside the cut and increases it on the baselines inside the cut similar to \cite{patil2016}. However, the power in Stokes $I$, when using all baselines in calibration, seems to decrease to smaller scales. Some decrease in power might occur when a diffuse sky-model is not included in the calibration and is calibrated away.

\item The discontinuity in the ratio of differential Stokes $I$ and $V$ power spectra for two calibration strategies appears at the location of calibration cut. We suggest that this effect is purely an artifact of the calibration cut. If the gains estimated during the calibration process using only a subset of baselines are erroneous, then the errors on gain estimates might transfer to smaller baselines, which are excluded in the calibration process. This enhances the excess noise on the excluded baselines (e.g. \citealt{barry2016,patil2016}). These errors on gains might occur due to incomplete sky-model and/or ionospheric scintillations.

\item We observe a `pitchfork' structure in the delay power spectrum of total polarized intensity ($\mathcal{P}$). The `pitchfork' structure appears due to bright sources ({Cas\,A} in our case) leaking from Stokes $I$ to $\mathcal{P}$ due to instrumental polarization. Most of the power on and around this structure disappears when {Cas\,A} is subtracted (using DD-calibration). The residual power after {Cas\,A} subtraction correlates strongly with the power before {Cas\,A} subtraction, suggesting inaccurate {Cas\,A} model and/or imperfect source subtraction during DD calibration. Subtraction of sources within the primary beam does not affect the `pitchfork'. 

\item Inclusion of short baselines in the calibration scheme suppresses the residual power around the `pitchfork' compared to the scheme where the short baselines are excluded from the calibration step. We expect that any unmodeled flux (diffuse) outside the primary beam gets absorbed in the gains when short baselines are used in the calibration step suppressing the power around the `pitchfork'. We show that the delay spectrum of $\mathcal{P}$ is a scaled down version of the Stokes $I$ delay spectrum. 

\item Ionospheric scintillations are dominant at low frequencies. The power spectrum of {Cas\,A} at small baselines (where {Cas\,A} can be treated as compact source) takes the form of a compact source corrupted by Kolmogorov-type turbulence. We observe extremely small ionospheric diffractive scales $r_{\text{diff}} \sim 400$ m towards {Cas\,A}. To our knowledge, these are the smallest scales ever measured at 60 MHz. The power spectrum of $\mathcal{P}$ in direction of {Cas\,A} fits very well with the Kolmogorov-type turbulence and appears to be a scaled down version of the Stokes $I$ power spectrum, which is another confirmation of the strong instrumental polarization leakage in LBA. 

\item Cross coherence between two residuals images of {Cas\,A}, when rotated to the phase center disappears on 5 minute intervals. This suggests that the residuals after {Cas\,A} subtraction are incoherent as expected for ionospheric scintillation noise, even though the coherent part of the source should be nearly constant in time for a source in the phase center. This also point towards strong ionospheric activity during observation. The ionosphere typically decorrelates on timescales of $\sim 10$s at lower frequencies. Solving for ionospheric effects in direction-dependent calibration step requires solutions intervals $\leq 10$s. This requires effectively calibrating each visibility snapshot and also requires higher SNR (to achieve better quality solutions) than afforded by the current LOFAR-LBA data.

\end{enumerate}

The contamination effects which we discussed in this work, although in part identified in LOFAR-HBA data at frequencies around 150 MHz, appear much stronger in LOFAR LBA data. This can in part be due to the small diffractive scale of the ionosphere, but also due to the calibration process and the incomplete sky-model. 
The level of these effects we have observed in our study is a clear indication that these and other far-field effects (such as scintillation of {Cas\,A}) pose much more severe concerns in current/upcoming CD experiments compared to the EoR experiments. These effects need to be accounted for before the thermal noise (or Stokes $V$ rms) level can be reached at frequencies relevant for 21-cm Cosmic Dawn observations. In upcoming CD experiments, such as with SKA-low, NENUFAR and LEDA, which will observe in the frequency range of $30$ to $80$ MHz, and will probe the same short baselines as studied here, these effects have to be mitigated to an accuracy of $\sim 0.01\%$ or be incoherent and below the thermal noise such that they average down in time in order to get a detection. This study will prove to be helpful in understanding the behavior of these contamination effects at low frequencies and mitigating them.

\section*{Acknowledgements}
BKG and LVEK acknowledge the financial support from NOVA cross-network grant. LOFAR, the Low Frequency Array designed and constructed by ASTRON, has facilities in several countries, that are owned by various parties (each with their own funding sources), and that are collectively operated by the International LOFAR Telescope (ILT) foundation under a joint scientific policy.




\bibliographystyle{mnras}
\bibliography{bibentries.bib} 

\begin{thebibliography}{}
\makeatletter
\relax
\def\mn@urlcharsother{\let\do\@makeother \do\$\do\&\do\#\do\^\do\_\do\%\do\~}
\def\mn@doi{\begingroup\mn@urlcharsother \@ifnextchar [ {\mn@doi@}
  {\mn@doi@[]}}
\def\mn@doi@[#1]#2{\def\@tempa{#1}\ifx\@tempa\@empty \href
  {http://dx.doi.org/#2} {doi:#2}\else \href {http://dx.doi.org/#2} {#1}\fi
  \endgroup}
\def\mn@eprint#1#2{\mn@eprint@#1:#2::\@nil}
\def\mn@eprint@arXiv#1{\href {http://arxiv.org/abs/#1} {{\tt arXiv:#1}}}
\def\mn@eprint@dblp#1{\href {http://dblp.uni-trier.de/rec/bibtex/#1.xml}
  {dblp:#1}}
\def\mn@eprint@#1:#2:#3:#4\@nil{\def\@tempa {#1}\def\@tempb {#2}\def\@tempc
  {#3}\ifx \@tempc \@empty \let \@tempc \@tempb \let \@tempb \@tempa \fi \ifx
  \@tempb \@empty \def\@tempb {arXiv}\fi \@ifundefined
  {mn@eprint@\@tempb}{\@tempb:\@tempc}{\expandafter \expandafter \csname
  mn@eprint@\@tempb\endcsname \expandafter{\@tempc}}}

\bibitem[\protect\citeauthoryear{{Asad} et~al.,}{{Asad}
  et~al.}{2015}]{asad2015}
{Asad} K.~M.~B.,  et~al., 2015, \mn@doi [\mnras] {10.1093/mnras/stv1107}, \href
  {http://adsabs.harvard.edu/abs/2015MNRAS.451.3709A} {451, 3709}

\bibitem[\protect\citeauthoryear{{Asad} et~al.,}{{Asad}
  et~al.}{2016}]{asad2016}
{Asad} K.~M.~B.,  et~al., 2016, \mn@doi [\mnras] {10.1093/mnras/stw1863}, \href
  {http://adsabs.harvard.edu/abs/2016MNRAS.462.4482A} {462, 4482}

\bibitem[\protect\citeauthoryear{{Asad}, {Koopmans}, {Jeli{\'c}}, {de Bruyn},
  {Pandey}  \& {Gehlot}}{{Asad} et~al.}{2017}]{asad2017}
{Asad} K.~M.~B.,  {Koopmans} L.~V.~E.,  {Jeli{\'c}} V.,  {de Bruyn} A.~G.,
  {Pandey} V.~N.,   {Gehlot} B.~K.,  2017, preprint, \href
  {http://adsabs.harvard.edu/abs/2017arXiv170600875A} {} (\mn@eprint {arXiv}
  {1706.00875})

\bibitem[\protect\citeauthoryear{{Asgekar} et~al.,}{{Asgekar}
  et~al.}{2013}]{asgekar2013}
{Asgekar} A.,  et~al., 2013, \mn@doi [\aap] {10.1051/0004-6361/201221001},
  \href {http://adsabs.harvard.edu/abs/2013A%26A...551L..11A} {551, L11}

\bibitem[\protect\citeauthoryear{{Aslanian}, {Dagkesamanskii}, {Kozhukhov},
  {Malumian}  \& {Sanamian}}{{Aslanian} et~al.}{1968}]{aslanian1968}
{Aslanian} A.~M.,  {Dagkesamanskii} R.~D.,  {Kozhukhov} V.~N.,  {Malumian}
  V.~G.,   {Sanamian} V.~A.,  1968, Astrofizika, \href
  {http://adsabs.harvard.edu/abs/1968Afz.....4..129A} {4}

\bibitem[\protect\citeauthoryear{{Baars}, {Genzel}, {Pauliny-Toth}  \&
  {Witzel}}{{Baars} et~al.}{1977}]{baars1977}
{Baars} J.~W.~M.,  {Genzel} R.,  {Pauliny-Toth} I.~I.~K.,   {Witzel} A.,  1977,
  \aap, \href {http://adsabs.harvard.edu/abs/1977A%26A....61...99B} {61, 99}

\bibitem[\protect\citeauthoryear{{Barning}}{{Barning}}{1963}]{barning1963}
{Barning} F.~J.~M.,  1963, \bain, \href
  {http://adsabs.harvard.edu/abs/1963BAN....17...22B} {17, 22}

\bibitem[\protect\citeauthoryear{{Barry}, {Hazelton}, {Sullivan}, {Morales}  \&
  {Pober}}{{Barry} et~al.}{2016}]{barry2016}
{Barry} N.,  {Hazelton} B.,  {Sullivan} I.,  {Morales} M.~F.,   {Pober} J.~C.,
  2016, \mn@doi [\mnras] {10.1093/mnras/stw1380}, \href
  {http://adsabs.harvard.edu/abs/2016MNRAS.461.3135B} {461, 3135}

\bibitem[\protect\citeauthoryear{{Becker} et~al.,}{{Becker}
  et~al.}{2001}]{becker2001}
{Becker} R.~H.,  et~al., 2001, \mn@doi [\aj] {10.1086/324231}, \href
  {http://adsabs.harvard.edu/abs/2001AJ....122.2850B} {122, 2850}

\bibitem[\protect\citeauthoryear{{Bernardi} et~al.,}{{Bernardi}
  et~al.}{2010}]{bernardi2010}
{Bernardi} G.,  et~al., 2010, \mn@doi [\aap] {10.1051/0004-6361/200913420},
  \href {http://adsabs.harvard.edu/abs/2010A%26A...522A..67B} {522, A67}

\bibitem[\protect\citeauthoryear{{Bolton}, {Becker}, {Wyithe}, {Haehnelt}  \&
  {Sargent}}{{Bolton} et~al.}{2010}]{bolton2010}
{Bolton} J.~S.,  {Becker} G.~D.,  {Wyithe} J.~S.~B.,  {Haehnelt} M.~G.,
  {Sargent} W.~L.~W.,  2010, \mn@doi [\mnras]
  {10.1111/j.1365-2966.2010.16701.x}, \href
  {http://adsabs.harvard.edu/abs/2010MNRAS.406..612B} {406, 612}

\bibitem[\protect\citeauthoryear{{Bowman} et~al.,}{{Bowman}
  et~al.}{2013}]{bowman2013}
{Bowman} J.~D.,  et~al., 2013, \mn@doi [\pasa] {10.1017/pas.2013.009}, \href
  {http://adsabs.harvard.edu/abs/2013PASA...30...31B} {30, e031}

\bibitem[\protect\citeauthoryear{Bregman}{Bregman}{2012}]{bregman2012}
Bregman J.,  2012, System design and wide-field imaging aspects of synthesis
  arrays with phased array stations: to the next generation of SKA system
  designers

\bibitem[\protect\citeauthoryear{{Brentjens} \& {de Bruyn}}{{Brentjens} \& {de
  Bruyn}}{2005}]{brentjens2005}
{Brentjens} M.~A.,  {de Bruyn} A.~G.,  2005, \mn@doi [\aap]
  {10.1051/0004-6361:20052990}, \href
  {http://adsabs.harvard.edu/abs/2005A%26A...441.1217B} {441, 1217}

\bibitem[\protect\citeauthoryear{{DeBoer} et~al.,}{{DeBoer}
  et~al.}{2017}]{deboer2017}
{DeBoer} D.~R.,  et~al., 2017, \mn@doi [\pasp]
  {10.1088/1538-3873/129/974/045001}, \href
  {http://adsabs.harvard.edu/abs/2017PASP..129d5001D} {129, 045001}

\bibitem[\protect\citeauthoryear{{Ewall-Wice} et~al.,}{{Ewall-Wice}
  et~al.}{2016}]{ewall-wice2016}
{Ewall-Wice} A.,  et~al., 2016, \mn@doi [\mnras] {10.1093/mnras/stw1022}, \href
  {http://adsabs.harvard.edu/abs/2016MNRAS.460.4320E} {460, 4320}

\bibitem[\protect\citeauthoryear{{Ewall-Wice}, {Dillon}, {Liu}  \&
  {Hewitt}}{{Ewall-Wice} et~al.}{2017}]{ewall-wice2017}
{Ewall-Wice} A.,  {Dillon} J.~S.,  {Liu} A.,   {Hewitt} J.,  2017, \mn@doi
  [\mnras] {10.1093/mnras/stx1221}, \href
  {http://adsabs.harvard.edu/abs/2017MNRAS.470.1849E} {470, 1849}

\bibitem[\protect\citeauthoryear{{Fan} et~al.,}{{Fan} et~al.}{2003}]{fan2003}
{Fan} X.,  et~al., 2003, \mn@doi [\aj] {10.1086/368246}, \href
  {http://adsabs.harvard.edu/abs/2003AJ....125.1649F} {125, 1649}

\bibitem[\protect\citeauthoryear{{Fan} et~al.,}{{Fan} et~al.}{2006}]{fan2006}
{Fan} X.,  et~al., 2006, \mn@doi [\aj] {10.1086/500296}, \href
  {http://adsabs.harvard.edu/abs/2006AJ....131.1203F} {131, 1203}

\bibitem[\protect\citeauthoryear{{Furlanetto}, {Oh}  \& {Briggs}}{{Furlanetto}
  et~al.}{2006}]{furlanetto2006}
{Furlanetto} S.~R.,  {Oh} S.~P.,   {Briggs} F.~H.,  2006, \mn@doi [\physrep]
  {10.1016/j.physrep.2006.08.002}, \href
  {http://adsabs.harvard.edu/abs/2006PhR...433..181F} {433, 181}

\bibitem[\protect\citeauthoryear{{Hamaker}, {Bregman}  \& {Sault}}{{Hamaker}
  et~al.}{1996}]{hamaker1996}
{Hamaker} J.~P.,  {Bregman} J.~D.,   {Sault} R.~J.,  1996, \aaps, \href
  {http://adsabs.harvard.edu/abs/1996A%26AS..117..137H} {117, 137}

\bibitem[\protect\citeauthoryear{{Hinshaw} et~al.,}{{Hinshaw}
  et~al.}{2013}]{hinshaw2013}
{Hinshaw} G.,  et~al., 2013, \mn@doi [\apjs] {10.1088/0067-0049/208/2/19},
  \href {http://adsabs.harvard.edu/abs/2013ApJS..208...19H} {208, 19}

\bibitem[\protect\citeauthoryear{{Jeli{\'c}} et~al.,}{{Jeli{\'c}}
  et~al.}{2015}]{jelic2015}
{Jeli{\'c}} V.,  et~al., 2015, \mn@doi [\aap] {10.1051/0004-6361/201526638},
  \href {http://adsabs.harvard.edu/abs/2015A%26A...583A.137J} {583, A137}

\bibitem[\protect\citeauthoryear{{Kassim} et~al.,}{{Kassim}
  et~al.}{2007}]{kassim2007}
{Kassim} N.~E.,  et~al., 2007, \mn@doi [\apjs] {10.1086/519022}, \href
  {http://adsabs.harvard.edu/abs/2007ApJS..172..686K} {172, 686}

\bibitem[\protect\citeauthoryear{{Kazemi} \& {Yatawatta}}{{Kazemi} \&
  {Yatawatta}}{2013}]{kazemi2013b}
{Kazemi} S.,  {Yatawatta} S.,  2013, \mn@doi [\mnras] {10.1093/mnras/stt1347},
  \href {http://adsabs.harvard.edu/abs/2013MNRAS.435..597K} {435, 597}

\bibitem[\protect\citeauthoryear{{Kazemi}, {Yatawatta}, {Zaroubi},
  {Lampropoulos}, {de Bruyn}, {Koopmans}  \& {Noordam}}{{Kazemi}
  et~al.}{2011}]{kazemi2011}
{Kazemi} S.,  {Yatawatta} S.,  {Zaroubi} S.,  {Lampropoulos} P.,  {de Bruyn}
  A.~G.,  {Koopmans} L.~V.~E.,   {Noordam} J.,  2011, \mn@doi [\mnras]
  {10.1111/j.1365-2966.2011.18506.x}, \href
  {http://adsabs.harvard.edu/abs/2011MNRAS.414.1656K} {414, 1656}

\bibitem[\protect\citeauthoryear{{Kazemi}, {Yatawatta}  \& {Zaroubi}}{{Kazemi}
  et~al.}{2013}]{kazemi2013a}
{Kazemi} S.,  {Yatawatta} S.,   {Zaroubi} S.,  2013, \mn@doi [\mnras]
  {10.1093/mnras/stt018}, \href
  {http://adsabs.harvard.edu/abs/2013MNRAS.430.1457K} {430, 1457}

\bibitem[\protect\citeauthoryear{{Kohn} et~al.,}{{Kohn}
  et~al.}{2016}]{kohn2016}
{Kohn} S.~A.,  et~al., 2016, \mn@doi [\apj] {10.3847/0004-637X/823/2/88}, \href
  {http://adsabs.harvard.edu/abs/2016ApJ...823...88K} {823, 88}

\bibitem[\protect\citeauthoryear{{Komatsu} et~al.,}{{Komatsu}
  et~al.}{2011}]{komatsu2011}
{Komatsu} E.,  et~al., 2011, \mn@doi [\apjs] {10.1088/0067-0049/192/2/18},
  \href {http://adsabs.harvard.edu/abs/2011ApJS..192...18K} {192, 18}

\bibitem[\protect\citeauthoryear{{Koopmans}}{{Koopmans}}{2010}]{koopmans2010}
{Koopmans} L.~V.~E.,  2010, \mn@doi [\apj] {10.1088/0004-637X/718/2/963}, \href
  {http://adsabs.harvard.edu/abs/2010ApJ...718..963K} {718, 963}

\bibitem[\protect\citeauthoryear{{Koopmans} et~al.,}{{Koopmans}
  et~al.}{2015}]{koopmans2015}
{Koopmans} L.,  et~al., 2015, Advancing Astrophysics with the Square Kilometre
  Array (AASKA14), \href {http://adsabs.harvard.edu/abs/2015aska.confE...1K}
  {p.~1}

\bibitem[\protect\citeauthoryear{{Lomb}}{{Lomb}}{1976}]{lomb1976}
{Lomb} N.~R.,  1976, \mn@doi [\apss] {10.1007/BF00648343}, \href
  {http://adsabs.harvard.edu/abs/1976Ap%26SS..39..447L} {39, 447}

\bibitem[\protect\citeauthoryear{{Madau}, {Meiksin}  \& {Rees}}{{Madau}
  et~al.}{1997}]{madau1997}
{Madau} P.,  {Meiksin} A.,   {Rees} M.~J.,  1997, \mn@doi [\apj]
  {10.1086/303549}, \href {http://adsabs.harvard.edu/abs/1997ApJ...475..429M}
  {475, 429}

\bibitem[\protect\citeauthoryear{{Mellema} et~al.,}{{Mellema}
  et~al.}{2013}]{mellema2013}
{Mellema} G.,  et~al., 2013, \mn@doi [Experimental Astronomy]
  {10.1007/s10686-013-9334-5}, \href
  {http://adsabs.harvard.edu/abs/2013ExA....36..235M} {36, 235}

\bibitem[\protect\citeauthoryear{{Mevius} et~al.,}{{Mevius}
  et~al.}{2016}]{mevius2016}
{Mevius} M.,  et~al., 2016, \mn@doi [Radio Science] {10.1002/2016RS006028},
  \href {http://adsabs.harvard.edu/abs/2016RaSc...51..927M} {51, 927}

\bibitem[\protect\citeauthoryear{{Mohan} \& {Rafferty}}{{Mohan} \&
  {Rafferty}}{2015}]{mohan2015}
{Mohan} N.,  {Rafferty} D.,  2015, {PyBDSF: Python Blob Detection and Source
  Finder}, Astrophysics Source Code Library (\mn@eprint {ascl} {1502.007})

\bibitem[\protect\citeauthoryear{{Offringa}, {de Bruyn}, {Biehl}, {Zaroubi},
  {Bernardi}  \& {Pandey}}{{Offringa} et~al.}{2010}]{offringa2010}
{Offringa} A.~R.,  {de Bruyn} A.~G.,  {Biehl} M.,  {Zaroubi} S.,  {Bernardi}
  G.,   {Pandey} V.~N.,  2010, \mn@doi [\mnras]
  {10.1111/j.1365-2966.2010.16471.x}, \href
  {http://adsabs.harvard.edu/abs/2010MNRAS.405..155O} {405, 155}

\bibitem[\protect\citeauthoryear{{Offringa}, {van de Gronde}  \&
  {Roerdink}}{{Offringa} et~al.}{2012}]{offringa2012}
{Offringa} A.~R.,  {van de Gronde} J.~J.,   {Roerdink} J.~B.~T.~M.,  2012,
  \mn@doi [\aap] {10.1051/0004-6361/201118497}, \href
  {http://adsabs.harvard.edu/abs/2012A%26A...539A..95O} {539, A95}

\bibitem[\protect\citeauthoryear{{Offringa} et~al.,}{{Offringa}
  et~al.}{2013a}]{offringa2013a}
{Offringa} A.~R.,  et~al., 2013a, \mn@doi [\mnras] {10.1093/mnras/stt1337},
  \href {http://adsabs.harvard.edu/abs/2013MNRAS.435..584O} {435, 584}

\bibitem[\protect\citeauthoryear{{Offringa} et~al.,}{{Offringa}
  et~al.}{2013b}]{offringa2013b}
{Offringa} A.~R.,  et~al., 2013b, \mn@doi [\aap] {10.1051/0004-6361/201220293},
  \href {http://adsabs.harvard.edu/abs/2013A%26A...549A..11O} {549, A11}

\bibitem[\protect\citeauthoryear{{Offringa} et~al.,}{{Offringa}
  et~al.}{2014}]{offringa2014}
{Offringa} A.~R.,  et~al., 2014, \mn@doi [\mnras] {10.1093/mnras/stu1368},
  \href {http://adsabs.harvard.edu/abs/2014MNRAS.444..606O} {444, 606}

\bibitem[\protect\citeauthoryear{{Ono} et~al.,}{{Ono} et~al.}{2012}]{ono2012}
{Ono} Y.,  et~al., 2012, \mn@doi [\apj] {10.1088/0004-637X/744/2/83}, \href
  {http://adsabs.harvard.edu/abs/2012ApJ...744...83O} {744, 83}

\bibitem[\protect\citeauthoryear{{Paciga} et~al.,}{{Paciga}
  et~al.}{2011}]{paciga2011}
{Paciga} G.,  et~al., 2011, \mn@doi [\mnras]
  {10.1111/j.1365-2966.2011.18208.x}, \href
  {http://adsabs.harvard.edu/abs/2011MNRAS.413.1174P} {413, 1174}

\bibitem[\protect\citeauthoryear{{Page} et~al.,}{{Page}
  et~al.}{2007}]{page2007}
{Page} L.,  et~al., 2007, \mn@doi [\apjs] {10.1086/513699}, \href
  {http://adsabs.harvard.edu/abs/2007ApJS..170..335P} {170, 335}

\bibitem[\protect\citeauthoryear{{Pandey}, {van Zwieten}, {de Bruyn}  \&
  {Nijboer}}{{Pandey} et~al.}{2009}]{pandey2009}
{Pandey} V.~N.,  {van Zwieten} J.~E.,  {de Bruyn} A.~G.,   {Nijboer} R.,  2009,
  in {Saikia} D.~J.,  {Green} D.~A.,  {Gupta} Y.,   {Venturi} T.,  eds,
  Astronomical Society of the Pacific Conference Series Vol. 407, The
  Low-Frequency Radio Universe. p.~384

\bibitem[\protect\citeauthoryear{{Parsons} \& {Backer}}{{Parsons} \&
  {Backer}}{2009}]{parsons2009}
{Parsons} A.~R.,  {Backer} D.~C.,  2009, \mn@doi [\aj]
  {10.1088/0004-6256/138/1/219}, \href
  {http://adsabs.harvard.edu/abs/2009AJ....138..219P} {138, 219}

\bibitem[\protect\citeauthoryear{{Parsons} et~al.,}{{Parsons}
  et~al.}{2010}]{parsons2010}
{Parsons} A.~R.,  et~al., 2010, \mn@doi [\aj] {10.1088/0004-6256/139/4/1468},
  \href {http://adsabs.harvard.edu/abs/2010AJ....139.1468P} {139, 1468}

\bibitem[\protect\citeauthoryear{{Parsons}, {Pober}, {Aguirre}, {Carilli},
  {Jacobs}  \& {Moore}}{{Parsons} et~al.}{2012}]{parsons2012}
{Parsons} A.~R.,  {Pober} J.~C.,  {Aguirre} J.~E.,  {Carilli} C.~L.,  {Jacobs}
  D.~C.,   {Moore} D.~F.,  2012, \mn@doi [\apj] {10.1088/0004-637X/756/2/165},
  \href {http://adsabs.harvard.edu/abs/2012ApJ...756..165P} {756, 165}

\bibitem[\protect\citeauthoryear{{Patil} et~al.,}{{Patil}
  et~al.}{2016}]{patil2016}
{Patil} A.~H.,  et~al., 2016, \mn@doi [\mnras] {10.1093/mnras/stw2277}, \href
  {http://adsabs.harvard.edu/abs/2016MNRAS.463.4317P} {463, 4317}

\bibitem[\protect\citeauthoryear{{Patil} et~al.,}{{Patil}
  et~al.}{2017}]{patil2017}
{Patil} A.~H.,  et~al., 2017, \mn@doi [\apj] {10.3847/1538-4357/aa63e7}, \href
  {http://adsabs.harvard.edu/abs/2017ApJ...838...65P} {838, 65}

\bibitem[\protect\citeauthoryear{{Pentericci} et~al.,}{{Pentericci}
  et~al.}{2011}]{pentericci2011}
{Pentericci} L.,  et~al., 2011, \mn@doi [\apj] {10.1088/0004-637X/743/2/132},
  \href {http://adsabs.harvard.edu/abs/2011ApJ...743..132P} {743, 132}

\bibitem[\protect\citeauthoryear{{Planck Collaboration} et~al.,}{{Planck
  Collaboration} et~al.}{2016}]{planck2016}
{Planck Collaboration} et~al., 2016, \mn@doi [\aap]
  {10.1051/0004-6361/201628897}, \href
  {http://adsabs.harvard.edu/abs/2016A%26A...596A.108P} {596, A108}

\bibitem[\protect\citeauthoryear{{Price} et~al.,}{{Price}
  et~al.}{2017}]{price2017}
{Price} D.~C.,  et~al., 2017, preprint, \href
  {http://adsabs.harvard.edu/abs/2017arXiv170909313P} {} (\mn@eprint {arXiv}
  {1709.09313})

\bibitem[\protect\citeauthoryear{{Pritchard} \& {Furlanetto}}{{Pritchard} \&
  {Furlanetto}}{2007}]{pritchard2007}
{Pritchard} J.~R.,  {Furlanetto} S.~R.,  2007, \mn@doi [\mnras]
  {10.1111/j.1365-2966.2007.11519.x}, \href
  {http://adsabs.harvard.edu/abs/2007MNRAS.376.1680P} {376, 1680}

\bibitem[\protect\citeauthoryear{{Pritchard} \& {Loeb}}{{Pritchard} \&
  {Loeb}}{2012}]{pritchard2012}
{Pritchard} J.~R.,  {Loeb} A.,  2012, \mn@doi [Reports on Progress in Physics]
  {10.1088/0034-4885/75/8/086901}, \href
  {http://adsabs.harvard.edu/abs/2012RPPh...75h6901P} {75, 086901}

\bibitem[\protect\citeauthoryear{{Rufenach}}{{Rufenach}}{1972}]{rufenach1972}
{Rufenach} C.~L.,  1972, \mn@doi [\jgr] {10.1029/JA077i025p04761}, \href
  {http://adsabs.harvard.edu/abs/1972JGR....77.4761R} {77, 4761}

\bibitem[\protect\citeauthoryear{{Schenker}, {Stark}, {Ellis}, {Robertson},
  {Dunlop}, {McLure}, {Kneib}  \& {Richard}}{{Schenker}
  et~al.}{2012}]{schenker2012}
{Schenker} M.~A.,  {Stark} D.~P.,  {Ellis} R.~S.,  {Robertson} B.~E.,  {Dunlop}
  J.~S.,  {McLure} R.~J.,  {Kneib} J.-P.,   {Richard} J.,  2012, \mn@doi [\apj]
  {10.1088/0004-637X/744/2/179}, \href
  {http://adsabs.harvard.edu/abs/2012ApJ...744..179S} {744, 179}

\bibitem[\protect\citeauthoryear{{Shaver}, {Windhorst}, {Madau}  \& {de
  Bruyn}}{{Shaver} et~al.}{1999}]{shaver1999}
{Shaver} P.~A.,  {Windhorst} R.~A.,  {Madau} P.,   {de Bruyn} A.~G.,  1999,
  \aap, \href {http://adsabs.harvard.edu/abs/1999A%26A...345..380S} {345, 380}

\bibitem[\protect\citeauthoryear{{Singleton}}{{Singleton}}{1974}]{singleton1974}
{Singleton} D.~G.,  1974, \mn@doi [Journal of Atmospheric and Terrestrial
  Physics] {10.1016/0021-9169(74)90071-3}, \href
  {http://adsabs.harvard.edu/abs/1974JATP...36..113S} {36, 113}

\bibitem[\protect\citeauthoryear{{Smirnov}}{{Smirnov}}{2011a}]{smirnov2011a}
{Smirnov} O.~M.,  2011a, \mn@doi [\aap] {10.1051/0004-6361/201016082}, \href
  {http://adsabs.harvard.edu/abs/2011A%26A...527A.106S} {527, A106}

\bibitem[\protect\citeauthoryear{{Smirnov}}{{Smirnov}}{2011b}]{smirnov2011b}
{Smirnov} O.~M.,  2011b, \mn@doi [\aap] {10.1051/0004-6361/201116434}, \href
  {http://adsabs.harvard.edu/abs/2011A%26A...527A.107S} {527, A107}

\bibitem[\protect\citeauthoryear{Stoica, Li  \& He}{Stoica
  et~al.}{2009}]{stoica2009}
Stoica P.,  Li J.,   He H.,  2009, \mn@doi [IEEE Transactions on Signal
  Processing] {10.1109/TSP.2008.2008973}, 57, 843

\bibitem[\protect\citeauthoryear{{Theuns}, {Schaye}, {Zaroubi}, {Kim},
  {Tzanavaris}  \& {Carswell}}{{Theuns} et~al.}{2002}]{theuns2002}
{Theuns} T.,  {Schaye} J.,  {Zaroubi} S.,  {Kim} T.-S.,  {Tzanavaris} P.,
  {Carswell} B.,  2002, \mn@doi [\apjl] {10.1086/339998}, \href
  {http://adsabs.harvard.edu/abs/2002ApJ...567L.103T} {567, L103}

\bibitem[\protect\citeauthoryear{{Thyagarajan} et~al.,}{{Thyagarajan}
  et~al.}{2015a}]{thyagarajan2015a}
{Thyagarajan} N.,  et~al., 2015a, \mn@doi [\apj] {10.1088/0004-637X/804/1/14},
  \href {http://adsabs.harvard.edu/abs/2015ApJ...804...14T} {804, 14}

\bibitem[\protect\citeauthoryear{{Thyagarajan} et~al.,}{{Thyagarajan}
  et~al.}{2015b}]{thyagarajan2015b}
{Thyagarajan} N.,  et~al., 2015b, \mn@doi [\apjl]
  {10.1088/2041-8205/807/2/L28}, \href
  {http://adsabs.harvard.edu/abs/2015ApJ...807L..28T} {807, L28}

\bibitem[\protect\citeauthoryear{{Tingay} et~al.,}{{Tingay}
  et~al.}{2013}]{tingay2013}
{Tingay} S.~J.,  et~al., 2013, \mn@doi [\pasa] {10.1017/pasa.2012.007}, \href
  {http://adsabs.harvard.edu/abs/2013PASA...30....7T} {30, e007}

\bibitem[\protect\citeauthoryear{{Trott} et~al.,}{{Trott}
  et~al.}{2016}]{trott2016}
{Trott} C.~M.,  et~al., 2016, \mn@doi [\apj] {10.3847/0004-637X/818/2/139},
  \href {http://adsabs.harvard.edu/abs/2016ApJ...818..139T} {818, 139}

\bibitem[\protect\citeauthoryear{{Vedantham} \& {Koopmans}}{{Vedantham} \&
  {Koopmans}}{2015}]{vedantham2015}
{Vedantham} H.~K.,  {Koopmans} L.~V.~E.,  2015, \mn@doi [\mnras]
  {10.1093/mnras/stv1594}, \href
  {http://adsabs.harvard.edu/abs/2015MNRAS.453..925V} {453, 925}

\bibitem[\protect\citeauthoryear{{Vedantham} \& {Koopmans}}{{Vedantham} \&
  {Koopmans}}{2016}]{vedantham2016}
{Vedantham} H.~K.,  {Koopmans} L.~V.~E.,  2016, \mn@doi [\mnras]
  {10.1093/mnras/stw443}, \href
  {http://adsabs.harvard.edu/abs/2016MNRAS.458.3099V} {458, 3099}

\bibitem[\protect\citeauthoryear{{Wang}}{{Wang}}{2013}]{wang2013}
{Wang} F.~Y.,  2013, \mn@doi [\aap] {10.1051/0004-6361/201321623}, \href
  {http://adsabs.harvard.edu/abs/2013A%26A...556A..90W} {556, A90}

\bibitem[\protect\citeauthoryear{{Yatawatta}}{{Yatawatta}}{2013}]{yatawatta2013}
{Yatawatta} S.,  2013, \mn@doi [Experimental Astronomy]
  {10.1007/s10686-012-9318-x}, \href
  {http://adsabs.harvard.edu/abs/2013ExA....35..469Y} {35, 469}

\bibitem[\protect\citeauthoryear{{Yatawatta}}{{Yatawatta}}{2015}]{yatawatta2015}
{Yatawatta} S.,  2015, \mn@doi [\mnras] {10.1093/mnras/stv596}, \href
  {http://adsabs.harvard.edu/abs/2015MNRAS.449.4506Y} {449, 4506}

\bibitem[\protect\citeauthoryear{Yatawatta}{Yatawatta}{2016}]{yatawatta2016}
Yatawatta S.,  2016, in 2016 24th European Signal Processing Conference
  (EUSIPCO). pp 265--269, \mn@doi{10.1109/EUSIPCO.2016.7760251}

\bibitem[\protect\citeauthoryear{{Zahn} et~al.,}{{Zahn}
  et~al.}{2012}]{zahn2012}
{Zahn} O.,  et~al., 2012, \mn@doi [\apj] {10.1088/0004-637X/756/1/65}, \href
  {http://adsabs.harvard.edu/abs/2012ApJ...756...65Z} {756, 65}

\bibitem[\protect\citeauthoryear{{Zarka}, {Girard}, {Tagger}  \&
  {Denis}}{{Zarka} et~al.}{2012}]{zarka2012}
{Zarka} P.,  {Girard} J.~N.,  {Tagger} M.,   {Denis} L.,  2012, in {Boissier}
  S.,  {de Laverny} P.,  {Nardetto} N.,  {Samadi} R.,  {Valls-Gabaud} D.,
  {Wozniak} H.,  eds, SF2A-2012: Proceedings of the Annual meeting of the
  French Society of Astronomy and Astrophysics. pp 687--694

\bibitem[\protect\citeauthoryear{{Zaroubi}}{{Zaroubi}}{2013}]{zaroubi2013}
{Zaroubi} S.,  2013, in {Wiklind} T.,  {Mobasher} B.,   {Bromm} V.,  eds,
  Astrophysics and Space Science Library Vol. 396, The First Galaxies. p.~45
  (\mn@eprint {arXiv} {1206.0267}), \mn@doi{10.1007/978-3-642-32362-1_2}

\bibitem[\protect\citeauthoryear{{Zheng}, {Wu}, {Johnston-Hollitt}, {Gu}  \&
  {Xu}}{{Zheng} et~al.}{2016}]{zheng2016}
{Zheng} Q.,  {Wu} X.-P.,  {Johnston-Hollitt} M.,  {Gu} J.-h.,   {Xu} H.,  2016,
  \mn@doi [\apj] {10.3847/0004-637X/832/2/190}, \href
  {http://adsabs.harvard.edu/abs/2016ApJ...832..190Z} {832, 190}

\bibitem[\protect\citeauthoryear{{van Haarlem} et~al.,}{{van Haarlem}
  et~al.}{2013}]{vanhaarlem2013}
{van Haarlem} M.~P.,  et~al., 2013, \mn@doi [\aap]
  {10.1051/0004-6361/201220873}, \href
  {http://adsabs.harvard.edu/abs/2013A%26A...556A...2V} {556, A2}

\bibitem[\protect\citeauthoryear{{van Straten}}{{van
  Straten}}{2009}]{vanstraten2009}
{van Straten} W.,  2009, \mn@doi [\apj] {10.1088/0004-637X/694/2/1413}, \href
  {http://adsabs.harvard.edu/abs/2009ApJ...694.1413V} {694, 1413}

\makeatother
\end{thebibliography}



\appendix

\section{Rotation Measure synthesis} 
\label{appendix:rmsyn}
The rotation of the polarization angle ($\chi$) of an electromagnetic wave, while propagating through magnetized plasma is called Faraday rotation. The value of $\chi$ depends on the frequency ($\nu$) of the wave, electron density ($n_e$) and magnetic field component parallel to the line of sight ($B_{\parallel}$). For a single Faraday screen, $\chi$ can be written as $\chi = {\chi}_{\circ} + \Phi \lambda^2$, where $\chi_{\circ}$ is the intrinsic polarization angle of the wave and $\Phi$ is the Faraday depth, which can be expressed as  
\begin{equation}
\dfrac{\Phi}{[\textrm{rad m}^{-2}]} = 0.81 \int_{\textrm{source}}^{\textrm{observer}} \dfrac{n_e}{[\textrm{cm}^{-3}]} \dfrac{B_{\parallel}}{[\mu\textrm{G}]} \dfrac{dl}{[\textrm{pc}]} .
\end{equation}		
The rotation measure is defined as the slope of $\chi(\lambda^2)$: 
\begin{equation}
\text{RM} = \dfrac{d\chi(\lambda^2)}{d\lambda^2} \text{ where, } \chi = \dfrac{1}{2} \text{tan}^{-1} \left( \dfrac{U}{Q} \right),
\end{equation}

and $Q$ and $U$ are the Stokes parameters. The RM synthesis technique \citep{brentjens2005} takes the advantage of the $\lambda^2$ dependence of the complex polarized emission ${\cal P}(\lambda^2) = Q(\lambda^2) + iU(\lambda^2)$. Using this, the complex Faraday dispersion function $F(\Phi)$ (which measures the Faraday rotation) can be defined as

\begin{equation}
\hspace{1cm}
F(\Phi) = R(\Phi) \ast \int_{-\infty}^{\infty} {\cal P}(\lambda^2) e^{-2i\Phi\lambda^2} d\lambda^2 , 
\end{equation}
\\
where $R(\Phi)$ is the Fourier transform of the wavelength sampling function $W(\lambda^2)$ and is known as the Rotation Measure Spread Function (RMSF). Rotation Measure (RM) synthesis can be used to distinguish between intrinsic and instrumental polarization by examining the polarized emission in RM-space. Before performing the RM-synthesis, the time varying ionospheric Faraday rotation has to be corrected. This correction is a global Faraday rotation correction (or de-rotation) with low time resolution ($\sim 15$ minutes in our case) and applies a single correction (corresponding to the phase center) to the entire field. It does not correct for any differential Faraday rotation with variations on shorter time scales and as a function of position. The de-rotation can be performed after the DI calibration. We use the RM estimates as a function of time from the GPS data (see e.g. figure \ref{fig:RMvaration}) to perform de-rotation using the BBS package. We have produced RM-cubes (we do not show any RM-cubes in the paper) before and after applying de-rotation. We used Stokes $Q$ and $U$ images produced with $200 \lambda$ imaging scheme (see Table \ref{tab:img_params}) for RM synthesis, because diffuse polarized emission is significant only on small baselines. We observed that before de-rotation, $\mathcal{P}$ in RM-space appears noise-like except at $\Phi = 0$, which is dominated by the instrumental polarization leakage. The de-rotation causes the emission due to polarization leakage at $\Phi = 0$ to move to some other Faraday depth whose value depends on the integrated RM over the duration of observation. Apart from this shift, $\mathcal{P}$ RM-cubes appear similar (noise-like) to RM-cubes before de-rotation.

Amount of depolarization due to time varying ionospheric RM can be estimated using RM values as a function of time ($t$). For time dependent RM (RM $(t)$), $\chi (t)$ can be written as
\begin{equation}
\chi(t) = \chi_{\circ} + \text{RM}(t) \times \lambda^2 \ , 
\end{equation}
Since $\mathcal{P} = Q + iU$, the variation in Stokes $Q$ and $U$ due to $\chi(t)$ is  
\begin{equation}
Q(t) = \mathcal{P}_{\circ} \cos \chi(t) \ \text{and} \ U(t) = \mathcal{P}_{\circ} \sin \chi(t) \ .
\end{equation}
The remaining total polarized intensity $|\mathcal{P}|$ (after ionospheric depolarization) after integrating over the observation time is given by
\begin{equation}
|\mathcal{P}| = \sqrt{\langle Q\rangle_t^2 + \langle Q\rangle_t^2}.
\end{equation}
Assuming $\chi_{\circ} = 0$ and $\mathcal{P}_{\circ} = 1$ gives fractional polarized intensity after the depolarization. We observed that time varying ionospheric RM produces depolarization of $\sim 75-80\%$ for 56-70 MHz frequency range. \citealt{jelic2015} observed bright polarized emission ($\sim 10$ K) in 3C196 field at 150 MHz using LOFAR-HBA observations. Assuming a spectral index of -2.55, we get polarized emission of $\sim 100$ K at 60 MHz. After taking ionospheric depolarization into account, we expect polarized emission of $\sim 20$ K at 60 MHz. Since, we do not observe any polarized emission in 3C196 field at LBA frequencies, it means either Galactic polarized emission at low frequencies is depolarized more than $5\%$ by intervening magneto-ionic medium because Faraday rotation scales as $\lambda^2$; or the differential Faraday rotation due to the ionosphere is significant, since we only correct for the phase center. Thus, a combination of both Galactic and ionospheric depolarization might be the cause of the absence of any polarized emission at low frequencies.

The resolution in Faraday depth space is $\delta\Phi = 2\sqrt{3}/(\lambda_{\text{min}}^2 - \lambda_{\text{max}}^2)$, which corresponds to the FWHM of the RMSF. For the frequency range 56-70 MHz, $\delta\Phi \approx 0.33 \ \text{rad/m}^2$, while the largest structure that can be resolved is only $\Delta\Phi_{\text{max}} = \pi/\lambda_{\text{min}}^2 \approx 0.17 \ \text{rad/m}^2$. Whereas at 150 MHz, $\delta\Phi \sim 1.75 \ \text{rad/m}^2$ and $\Delta\Phi_{\text{max}} \sim 1.15 \ \text{rad/m}^2$ which is almost an order of magnitude larger compared to lower frequencies. It is possible that the polarized structures observed in 3C196 field at 150 MHz are Faraday thick ($\lambda^2\Delta\Phi >> 1$) 
at lower frequencies and therefore cannot be observed in LBA. This is similar to Faraday thick structures in 3C196 field at 150 MHz which are not observed with LOFAR-HBA but have been detected at 350 MHz with WSRT (see section 6 in \citealt{jelic2015}).



\bsp	
\label{lastpage}
\end{document}